\documentclass[journal,10pt]{IEEEtran}
\usepackage{algorithm} 
\makeatletter
\renewcommand\fs@ruled{%
  \def\@fs@cfont{\bfseries}%
  \let\@fs@capt\floatc@ruled
  \def\@fs@pre{{\color{black}\hrule height.8pt depth0pt}\kern2pt}%
  \def\@fs@post{\kern2pt{\color{black}\hrule}\relax}%
  \def\@fs@mid{\kern2pt{\color{black}\hrule}\kern2pt}%
  \let\@fs@iftopcapt\iftrue}
\makeatother
\usepackage{algorithmic} 
\usepackage{multirow} 
\usepackage{amsmath}
\usepackage{xcolor}
\usepackage{amsmath}
\usepackage{stfloats} 
\usepackage{amssymb}
\usepackage{amsmath}
\usepackage{cite}
\usepackage{hyperref}
\usepackage{url}
\usepackage{xcolor}
\usepackage{cite,graphicx,amsmath,amssymb}
\usepackage{fancyhdr}
\usepackage{mdwmath}
\usepackage{mdwtab}
\usepackage{caption}
\usepackage{amsthm}

\usepackage{setspace}
\usepackage{graphicx}

\usepackage{enumitem}
\usepackage{tabularx}

\usepackage{subcaption}

\newtheorem{remark}{Remark}
\newtheorem{theorem}{Theorem}

\newtheorem{lemma}{Lemma}

\newtheorem{corollary}{Corollary}

\makeatletter
\def\ScaleIfNeeded{%
\ifdim\Gin@nat@width>\linewidth \linewidth \else \Gin@nat@width
\fi } \makeatother

\begin{document}

\title{Distributed Trajectory Planning and Resource Allocation for Dynamic Multi-UAV Collaborative Computing}

\author{
Tiankui~Zhang,~\IEEEmembership{Senior Member,~IEEE}, 
Wenlong~Xu,
Tianyi~Shi,~\IEEEmembership{Graduate Student Member,~IEEE},
Xiaoxia~Xu,~\IEEEmembership{Member, IEEE},
Arumugam~Nallanathan,~\IEEEmembership{Fellow, IEEE}

\thanks{
This work is supported by National Natural Science Foundation of China under Grant 62371068.
}
\thanks{
Tiankui~Zhang, Wenlong~Xu and Tianyi~Shi are with the School of Information and Communication Engineering, Beijing University of Posts and Telecommunications, Beijing 100876, China (e-mail: zhangtiankui@bupt.edu.cn, Evan\_xwl@163.com, shi\_tianyi@bupt.edu.cn).
}
\thanks{
Xiaoxia Xu is with the School of Electronic Engineering and Computer Science, Queen Mary University of London, E1 4NS London, U.K. (e-mail: x.xiaoxia@qmul.ac.uk).
}
\thanks{A. Nallanathan is with the School of Electronic Engineering and Computer Science, Queen Mary University of London, London and also with the Department of Electronic Engineering, Kyung Hee University, Yongin-si, Gyeonggi-do 17104, Korea (email: a.nallanathan@qmul.ac.uk).}
\thanks{
Copyright (c) 2026 IEEE. Personal use of this material is permitted.
However, permission to use this material for any other purposes must be obtained
from the IEEE by sending a request to pubs-permissions@ieee.org.
}
}

\maketitle

\begin{abstract}
This paper investigates a multiple uncrewed aerial vehicles (UAVs)-enabled distributed mobile edge computing (MEC) framework, where the set of collaborative UAVs dynamically varies over time due to their energy states and service loads.
The joint optimization of trajectory planning and resource allocation is formulated as a Stackelberg game, where UAVs and mobile terminals (MTs) are modeled as leaders and followers, respectively.  
UAVs aim to maximize their benefits by balancing executed workload, energy cost, and resource allocation revenue, while the MTs seek to minimize their total overhead, composed of computing delay and resource costs, through offloading and resource-request decisions in response to the UAV-side service decisions.
A hierarchical joint optimization algorithm is developed within a multi-agent deep reinforcement learning (MADRL) framework to coordinate UAVs and MTs in a distributed manner.
At the leader level, UAVs jointly determine their trajectories, task migration ratios, MT-UAV association, and unit computing resource pricing. 
For distributed decision making, each UAV is modeled as an agent in a partially observable Markov decision process, and the agents are jointly trained via multi-agent proximal policy optimization (MAPPO) under the centralized-training-and-decentralized-execution paradigm.
At the follower level, MTs determine their optimal task offloading ratios and requested computing resources using a two-stage iterative algorithm. 
Simulation results demonstrate that the proposed algorithm achieves stable convergence under the dynamic UAV participation. 
Compared to the no-collaboration benchmark, it improves UAV efficiency by 18.58\% through effective balancing of computing workloads via inter-UAV task migration and reduces average MT overhead by 33.77\% over the fully offloading scheme. 
It also outperforms other benchmarks under varying network scales and capabilities by jointly optimizing UAV operations and resource utilization.

\end{abstract}

\providecommand{\keywords}[1]{\textbf{\textit{Index terms---}} #1}


\section{Introduction}
With the rapid development of mobile Internet and the wide application of various computing applications, mobile networks are facing rapidly growing demands for computing resources.
Mobile edge computing (MEC) reduces computing latency and network traffic pressure by deploying computing resources to the network edge closer to the mobile terminals (MTs)~\cite{mec_intro1,mec_intro2,mec_intro3}. 
In the absence of fixed network infrastructure, such as in remote field environments or emergency rescue scenarios, MEC systems can be flexibly deployed with the aid of uncrewed aerial vehicles (UAVs)~\cite{uav_mec_intro2,uav_mec_intro3}. 
As agile aerial platforms, UAVs can improve the quality of ground communication links and expand network coverage by establishing Line-of-Sight (LoS) links. 
Owing to these merits, they have received considerable attention in applications such as data acquisition, spectrum sensing, and emergency communication \cite{linxin,10679984}. 
These features make UAV-enabled MEC particularly attractive for on-demand computing services in infrastructure-limited scenarios.
\par

Traditional MEC systems adopt a centralized architecture, where MEC servers deployed in different geographical locations are typically managed by a centralized controller. This setup enables efficient resource management of MEC systems and collaborative computing among MEC servers. 
However, in such architecture, both MEC servers and MTs need to frequently send status information to the centralized controller to maintain service continuity \cite{9652164,10325641,centralized_shortcomings_1}. 
Hence, when directly applied to UAV-enabled MEC systems, this architecture faces two key challenges. 
On the one hand, the communication overhead and resource management complexity of the centralized controller increase significantly with network size. 
On the other hand, the high mobility of UAVs leads to frequent changes in network topology, which further complicates centralized management and control \cite{10123387,centralized_shortcomings_2}. 
These limitations hinder the scalability of UAV networks and confine flexible UAV deployment. 
As a result, UAV-enabled MEC leads to a highly dynamic service environment, where the evolving topology and resource availability make centralized coordination increasingly difficult.

\subsection{Related Work}
Existing research on UAV-enabled MEC systems can be broadly classified into three categories, i.e., single UAV-enabled MEC, centralized multi-UAV MEC, and distributed multi-UAV MEC, as analyzed as follows.

\subsubsection{Single UAV-enabled MEC}
The related work of single UAV-enabled MEC systems mainly focuses on resource allocation and trajectory planning \cite{9951396,reviewer1,10706975,10716541,10707659}. 
Wang et al. considered the MEC system composed of base stations and a UAV with complex channel environments in cities and jointly optimized computing resources, communication bandwidth, computing offloading, and UAV trajectory to maximize computational efficiency \cite{9951396}. 
Joint trajectory and resource optimization has also been investigated in UAV-assisted heterogeneous edge computing networks with D2D support~\cite{reviewer1}.
In order to minimize the weighted completion time and propulsion energy consumption of rotary-wing UAV, double deep Q-learning (DDQN) is used to optimize the three-dimensional trajectory of the UAV while maintaining the stability of the UAV ground terminal link \cite{10706975}. 
Hu et al. designed a layered aerial computing framework consisting of MTs, the UAV, and high-altitude platforms, maximizing model resource utilization and task scheduling, and proposed a trajectory optimization and task offloading algorithm based on the actor-critic approach \cite{10716541}. 
Wei et al. studied task partition and offloading between IoT devices and the UAV to improve training efficiency and reduce UAV energy consumption \cite{10707659}.
However, single-UAV MEC systems are inherently limited by restricted coverage, constrained onboard computing resources, and low service robustness. A single UAV cannot simultaneously approach all connected MTs, and its service capability is vulnerable to failure or interference. 
These limitations motivate the study of multi-UAV-enabled MEC systems, where service capability and system flexibility can be further enhanced through cooperation among multiple UAVs.

\subsubsection{Centralized Multi-UAV MEC}
Some studies that consider MEC systems with multiple UAVs~\cite{10711854,10713504,10705107} have each UAV independently responsible for its coverage area, ignoring the coordination between UAVs, which may lead to uneven offload distribution. 
To leverage the mutual advantages among multiple UAVs, considerable studies~\cite{10673993,10716389,10678860} have been conducted to further enhance the system performance.
In~\cite{10673993}, the author designed an aerial MEC architecture where multiple UAVs aggregate data onto a high-altitude platform and minimized system latency and energy consumption by jointly optimizing container caching decisions and UAV trajectories. 
In the case of dynamic task generation, Li et al. optimized task scheduling and UAV trajectories to maximize coverage and reduce energy consumption via  a multi-agent deep deterministic policy gradient (MADDPG)-based algorithm \cite{10716389}. 
Liu et al. explored the scenario of combining semantic communication with MEC, aiming to minimize task completion time and improve semantic spectrum efficiency \cite{10678860}. 
Nevertheless, the above studies do not take the dynamic environments of ground MTs' location movement into consideration. 
To address the challenges in dynamic environments, deep reinforcement learning (DRL) is widely used for trajectory and resource allocation optimization in UAV-enabled MEC systems \cite{10462233,reviewer2,10417719,comparison1,comparison2}. 
Liu et al. used deep Q-learning to optimize UAV positioning and resource allocation under dynamic interference \cite{10462233}.
Online service selection and task offloading in UAV-assisted MEC have also been investigated via deep reinforcement learning \cite{reviewer2}.
Considering both the random tasks and time-varying channels of MTs, the authors of \cite{10417719} proposed a contract incentive strategy based on DQN to maximize the long-term utility of all hotspots while maintaining energy stability. 
These works typically assume a central controller with global knowledge to conduct the optimization. Although some of them consider dynamic user movement, random task generation, or time-varying channels, the UAV-side decision process is still mainly handled in a centralized manner and the available UAV set is usually treated as fixed during optimization.
Although centralized approaches can enhance system efficiency, they face challenges such as high training complexity, limited scalability, and vulnerability to single points of failure in large-scale deployments, motivating the development of distributed architectures.

\subsubsection{Distributed Multi-UAV MEC}
Distributed multi-UAV MEC architectures have emerged as a promising solution to the limitations of centralized systems. 
By decentralizing decision-making, these frameworks reduce communication overhead, distribute computation loads on individual nodes, and improve system scalability~\cite{9162899}.  
Noor et al. constructed a vehicle networking system assisted by UAVs and proposed a distributed offloading and resource allocation algorithm through value-based iteration in reinforcement learning that minimizes the computational and communication overhead of vehicle systems \cite{9793590}.
Under the air-ground collaborative MEC architecture, Ye et al. proposed a low-complexity distributed DQN algorithm using the deterministic state transition characteristics of UAV-enabled MEC systems \cite{9299721}. 
Considering the limited channel capacity and system computing resources between UAVs and MTs, the author established a distributed DNN that improves training efficiency by simultaneously training multiple DNNs while minimizing system delay and energy consumption weighting \cite{9162899}. 
Li et al. developed an MT privacy and security computing offloading method based on distributed federated learning, which maximizes the normalized weighted sum of task response delay and MT energy consumption \cite{10665983}. 
Kim et al. proposed a collaborative MADDPG algorithm that introduces graph attention networks into DRL and designs multiple types of action variables, improving the system performance in a distributed manner \cite{10643151}.
These studies have demonstrated the potential of distributed offloading, trajectory control, and resource scheduling in multi-UAV MEC systems. 
However, existing distributed multi-UAV MEC studies usually make decisions over a static set of UAV agents, while network dynamics are mainly reflected in MT mobility, stochastic task arrivals, channel fluctuations, or workload variation. The case where the UAV collaboration set itself changes over time has received less attention.  Such UAV-side availability dynamics alter the service topology, available computing resources, and feasible coordination space among UAVs and MTs. Accordingly, how to jointly coordinate topology-dependent association and migration decisions with UAV-side pricing and MT-side responses under such dynamics remains insufficiently~explored.

\vspace{-8pt}
\subsection{Motivation and Contribution}
Although substantial research has been devoted to improving system performance and addressing practical deployment challenges, distributed multi-UAV MEC under dynamic UAV participation remains insufficiently studied. 
Existing dynamic MEC studies commonly focus on time-varying MT locations, stochastic task arrivals, fluctuating wireless channels, or changing service demands, while the available UAV collaboration set and the network scale are usually assumed to remain fixed during optimization.
In many practical missions, the set of available UAVs may vary over time due to energy states and workload conditions, which means that UAVs may join or leave the collaborative computing network and thus lead to a time-varying UAV-side service topology and decision space.
Effectively harnessing UAVs' inherent mobility and flexible deployment capabilities is crucial to fully realizing their potential in MEC scenarios. 
Under such UAV-side availability dynamics, the trajectory planning, MT-UAV association, inter-UAV task migration, and resource provisioning decisions become more tightly coupled and must adapt to the current service topology. 
To address this issue, this paper investigates a distributed MEC system that accommodates changes in ground MT locations and fluctuating service demands, where the available UAV set varies over time. 
Within this architecture, a trajectory planning and resource allocation problem is formulated based on a Stackelberg game, where UAVs and MTs act as leaders and followers, respectively. 
Multi-agent deep reinforcement learning (MADRL) is employed to enable UAVs to learn fast adaptive service decisions under the time-varying network scale and service topology, thus significantly reducing system-level control overhead. The main contributions of this article are summarized as follows.

\begin{itemize}[leftmargin=*]
  \item We propose a distributed multi-UAV-enabled MEC framework under dynamic UAV participation, where the available UAV set varies over time according to system conditions. The system model captures partial task offloading from MTs to serving UAVs, inter-UAV task migration for load balancing, time-varying MT locations, and stochastic task generation. Under the resulting time-varying service topology and decision space, we formulate the distributed joint trajectory planning and resource allocation problem within a Stackelberg game framework, where UAVs and MTs act as leaders and followers, respectively.

  \item We develop a hierarchical MADRL-based solution to enable distributed coordination between UAVs and MTs. At the leader level, each UAV is modeled as an agent under partial observability and trained via MAPPO under the centralized training and decentralized execution paradigm to optimize trajectories, task migration ratio, user association, and resource pricing. At the follower level, given the leaders' decisions, MTs update task offloading ratios and purchased computing resources via a two-stage iterative procedure, where the update rules are derived using the Lagrange dual method and Karush-Kuhn-Tucker (KKT) conditions.

  \item Simulation results validate the effectiveness of the proposed framework. The proposed algorithm exhibits good convergence behavior under time-varying UAV availability and achieves lower MT overhead and higher UAV-side benefits compared with benchmark methods under different computing capabilities, network scales, and bandwidth settings. 
\end{itemize}

\vspace{-10pt}
\subsection{Organization}
The rest of this paper is organized as follows. In Section~\ref{sec:sys}, we introduce the system model and formulate the corresponding Stackelberg game. A distributed hierarchical optimization algorithm is proposed in Section~\ref{sec:alg}. The performance of the proposed algorithm is evaluated by the simulation in Section~\ref{sec:simluation}, while the paper concludes in Section~\ref{sec:conclusion}.

\section{System Model}\label{sec:sys}
\begin{figure}[!t]
\centering
\includegraphics[width= 3.5in]{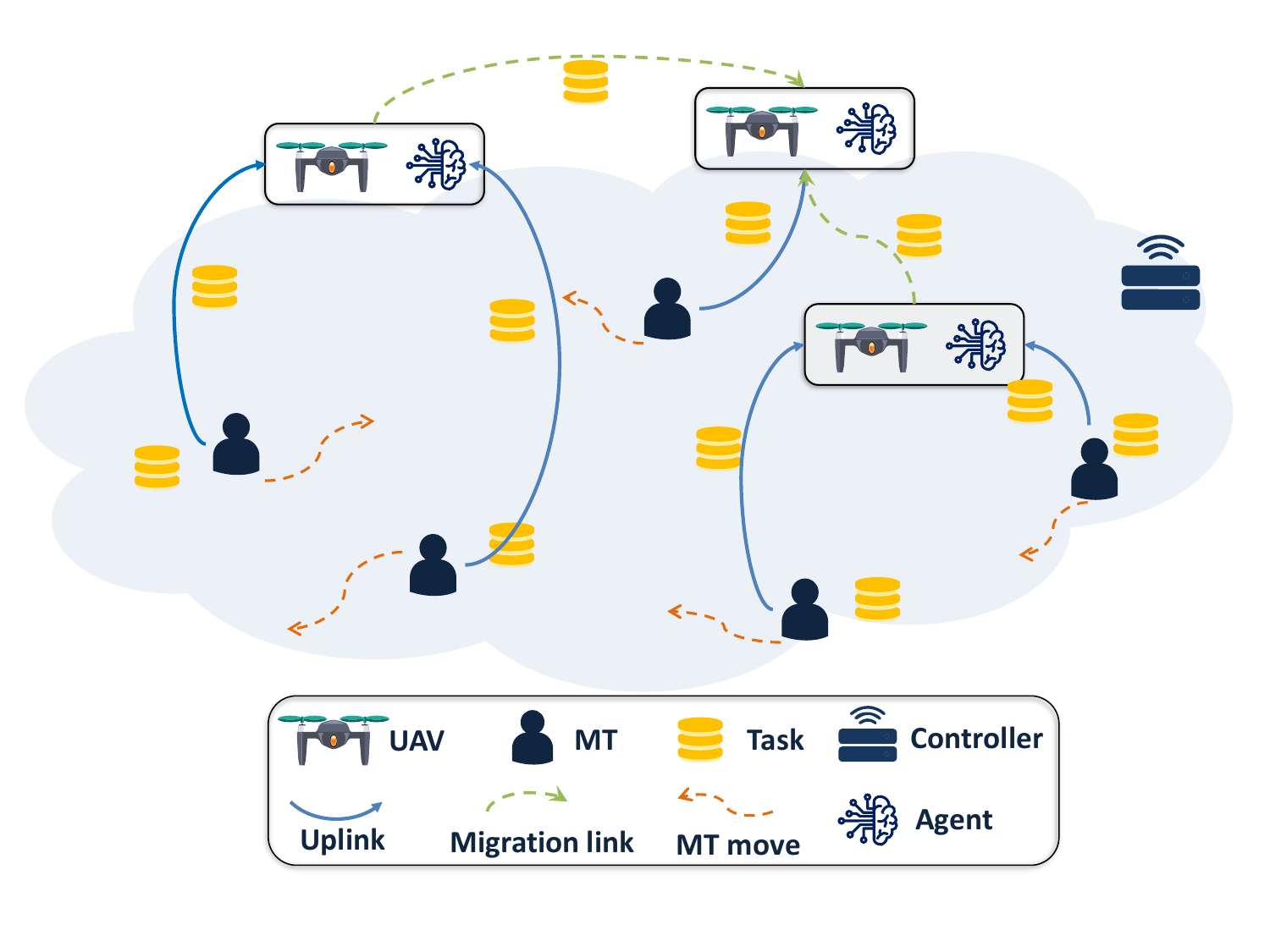}
\caption{Dynamic distributed UAV-enabled MEC system model.}
\label{sys}
\end{figure}

A dynamic distributed UAV-enabled MEC system is considered for industrial private networks, where multiple UAVs act as aerial MEC nodes to provide elastic computing and coverage enhancement for ground MTs, as illustrated in Fig.~\ref{sys}.
Using trajectory discretization, the considered horizon is partitioned into $T$ slots indexed by $t \in \mathcal{T} \triangleq \{1,2,\ldots,T\}$, each with an adaptive duration $\tau_t$ upper bounded by $\delta$, where $\delta$ denotes the maximum allowable slot length determined by the system design, and tasks generated in one slot are assumed to be completed within the same slot. The quasi-static approximation is adopted, indicating that the distance between each UAV and each MT is treated as constant when evaluating within a slot.
Let $\mathcal{K} \triangleq \{1,2,\ldots,K\}$ denote the MT set. Due to dynamic participation, the set of available UAVs at slot $t$ is $\mathcal{M}(t)$ with $|\mathcal{M}(t)|=M(t)$, where $M(t)$ can vary over time, and is viewed in this work as an exogenous time-varying system condition. 
At each slot, MTs generate computation tasks and can partially offload them to associated UAVs. Accordingly, the cooperative decision space varies over time with the current UAV availability. To cope with load imbalance, UAVs can further forward a portion of offloaded tasks to nearby UAVs for cooperative computing.
The main symbols defined in the system model are described in Table~\ref{tab1}.

\vspace{-10pt}

\begin{table}[t!]
\centering
\caption{Notation Descriptions}
\label{tab1}
\footnotesize
\resizebox{\linewidth}{!}{
\begin{tabular}{|l|l|}\hline
Notation & Description\\\hline
$\mathcal{T}$,$T$ & set and number of time slots \\ \hline
${\tau_t}$ & duration of time slot $t$\\ \hline
$\delta$ & maximum slot duration\\ \hline
$\mathcal{K}$, $K$ & set and number of MTs\\ \hline
$M(t)$ & number of active UAVs in slot $t$\\ \hline
${\mathbf{q}_m}(t)$ & horizontal coordinates of UAV $m$ in slot $t$\\ \hline
${\mathbf{w}_k}(t)$ & horizontal coordinates of MT $k$ in slot $t$\\ \hline
${v_m}(t)$ & flight speed of UAV $m$ in slot $t$\\ \hline
${\alpha_m}(t)$ & heading angle of UAV $m$ in slot $t$\\ \hline
${v_k}(t)$ & moving speed of MT $k$ in slot $t$\\ \hline
${\alpha_k}(t)$ & moving direction of MT $k$ in slot $t$\\ \hline
${J_k}(t)$ & task input data size (bits) of MT $k$ in slot $t$\\ \hline
${C_k}$ & required CPU cycles per bit for MT $k$'s task\\ \hline
${g_{m,m'}}(t)$ & channel power gain between UAV $m$ and UAV $m'$\\ \hline
${d_{m,m'}}(t)$ & distance between UAV $m$ and UAV $m'$\\ \hline
${d_{k,m}}(t)$ & distance between MT $k$ and UAV $m$\\ \hline
$\gamma_k(t)$ & offloading ratio of MT $k$\\ \hline
${z_{k,m}}(t)$ & indicator of MT association \\ \hline
${p_m}(t)$ & propulsion power of UAV $m$ in slot $t$\\ \hline
${r_{k,m}}(t)$ & uplink transmission rate from MT $k$ to UAV $m$\\ \hline 
${r_{m,m'}}(t)$ & transmission rate between UAV $m$ and UAV $m'$\\ \hline
${\rho_{k,m,m'}}(t)$ & inter-UAV task offloading ratio\\ \hline
${f_{k,m}}(t)$ & CPU allocation from UAV $m$ to MT $k$ in slot $t$\\ \hline
$b_m(t)$ & unit price of UAV $m$\\ \hline
\end{tabular}}
\end{table}

\subsection{UAV and MT Movement Models}
The flight altitude of all UAVs is fixed at a constant value $H$. 
In a Cartesian coordinate system, the horizontal coordinates of UAV $m$ in time slot $t$ can be represented as ${q_m}(t) = \left[ {{x_m}\left( t \right),{y_m}\left( t \right)} \right]$, and the horizontal starting point of UAV $m$ is defined as $q_m^{{\rm{ini}}}$. 
Given the flight speed ${v_m}\left( t \right) \in \left[ {0,{V_{\max }}} \right]$ and direction ${\alpha _m}\left( t \right)$, where ${V_{\max }}$ is the maximum UAV speed (m/s), the UAV position is updated as
\begin{align}
{x_m}\left( t \right) &= {x_m}\left( {t - 1} \right) + {v_m}\left( t \right)\cos \left( {{\alpha _m}\left( t \right)} \right){\tau _t}, \\
{y_m}\left( t \right) &= {y_m}\left( {t - 1} \right) + {v_m}\left( t \right)\sin \left( {{\alpha _m}\left( t \right)} \right){\tau _t}.
\end{align}

The horizontal coordinate of MT $k$ in time slot $t$ is defined as ${w_k}\left( t \right) = \left[ {{x_k}\left( t \right),{y_k}\left( t \right)} \right]$. 
Following~\cite{gauss_markov}, we adopt a Gauss--Markov mobility model for MT movements, which provides a tractable way to characterize the temporally correlated movement of MTs across adjacent slots. Specifically, the speed ${v_k}\left( t \right)$ (m/s) and moving direction ${\alpha _k}\left( t \right)$ (rad) are updated as
\begin{align}
{v_k}\left( t \right) &= {c_1}{v_k}\left( {t - 1} \right) + (1 - {c_1})\bar v + \sqrt {1 - c_1^2}\,{\Phi _k}, \\
{\alpha _k}\left( t \right) &= {c_2}{\alpha _k}\left( {t - 1} \right) + (1 - {c_2})\bar \alpha  + \sqrt {1 - c_2^2}\,{\Psi _k},
\end{align}
where ${c_1}\in[0,1]$ and ${c_2}\in[0,1]$ are correlation coefficients, $\bar v$ and $\bar \alpha$ are the mean speed and mean direction, respectively, and ${\Phi _k}$ and ${\Psi _k}$ are independent Gaussian random variables capturing random perturbations.
Accordingly, the MT position is updated as
\begin{align}
{x_k}\left( t \right) &= {x_k}\left( {t - 1} \right) + {v_k}\left( t \right)\cos \left( {{\alpha _k}\left( t \right)} \right){\tau _t}, \\
{y_k}\left( t \right) &= {y_k}\left( {t - 1} \right) + {v_k}\left( t \right)\sin \left( {{\alpha _k}\left( t \right)} \right){\tau _t}.
\end{align}

\subsection{Communication Model}
In practical application scenarios, due to the presence of obstacles such as buildings and mountains, the actual path loss between UAVs and MTs is determined by the probabilities of both LoS and Non-Line-of-Sight (NLoS) links. The probability that the link between MT $k$ and UAV $m$ in time slot $t$ is a LoS link is given by
\begin{align}
p_{k,m}^{{\rm{LoS}}}\left( t \right) = \frac{1}{{1 + {\eta _a}\exp \left( { - {\eta _b}\left( {{\theta _{k,m}}(t) - {\eta _a}} \right)} \right)}},
\end{align}
where ${\eta_a}$ and ${\eta _b}$ are constants related to the propagation environment. ${\theta _{k,m}}(t) = \arcsin \left( \frac{H}{{{d_{k,m}}\left( t \right)}} \right)$ is the elevation angle, and ${d_{k,m}}\left( t \right) = \sqrt {{H^2} + \left\| {{w_k}\left( t \right) - {q_m}\left( t \right)} \right\|_2^2} $ represents the Euclidean distance between MT $k$ and UAV $m$.
Accordingly, the probability that the link is an NLoS link is $p_{k,m}^{{\rm{NLoS}}}\left( t \right) = 1 - p_{k,m}^{{\rm{LoS}}}\left( t \right)$.
The free-space path loss between MT $k$ and UAV $m$ is modeled as
\begin{align}
L_{k,m}^{{\rm FS}}(t)=20\log_{10}\!\big(d_{k,m}(t)\big)+20\log_{10}\!\left(\frac{4\pi f_c}{v_c}\right),
\end{align}
where $f_c$ is the carrier frequency and $v_c$ is the speed of light.
Therefore, the average path loss is then given by
\begin{align}
L_{k,m}(t)=L_{k,m}^{{\rm FS}}(t)+p_{k,m}^{{\rm LoS}}(t)\eta_{{\rm LoS}}+p_{k,m}^{{\rm NLoS}}(t)\eta_{{\rm NLoS}},
\end{align}
where $\eta_{{\rm LoS}}$ and $\eta_{{\rm NLoS}}$ denote the additional loss factors for LoS and NLoS links, respectively.
\par

We consider an orthogonal frequency division multiple access (OFDMA)-based uplink for task data transmission. For tractability, the total bandwidth is equally partitioned among MTs, and the bandwidth allocated to each MT is denoted by $B_K$.
This assumption simplifies the uplink transmission model so that the proposed Stackelberg game can more clearly characterize the interaction between UAV pricing and service decisions and the MTs' offloading responses.
Thus, the uplink transmission rate from MT $k$ to UAV $m$ in slot $t$ is
\begin{align}
r_{k,m}(t)=B_K\log_2\!\left(1+\frac{P_K}{\sigma^2\,10^{L_{k,m}(t)/10}}\right),
\end{align}
where $\sigma^2$ is the noise power and $P_K$ is the MT transmit power.

Since air-to-air links are typically LoS due to the absence of obstacles in inter-UAV communications, the channel power gain between UAV $m$ and UAV $m'$ in slot $t$ is modeled as
\begin{align}
g_{m,m'}(t)=\beta_0 d_{m,m'}^{-2}(t)=\frac{\beta_0}{\left\|q_m(t)-q_{m'}(t)\right\|_2^2},
\end{align}
where $\beta_0$ is the channel power gain at the reference distance of 1 meter, and
$d_{m,m'}(t)=\left\|q_m(t)-q_{m'}(t)\right\|_2$ is the distance between UAV $m$ and UAV $m'$.
To ensure flight safety, we impose the minimum separation constraint $d_{m,m'}(t)\ge d^{\min}$.
Accordingly, the task transmission rate between UAV $m$ and UAV $m'$ is
\begin{align}
r_{m,m'}(t)=B_M\log_2\!\left(1+\frac{P_M g_{m,m'}(t)}{\sigma^2}\right),\quad \forall m\neq m'\in \mathcal{M}(t),
\end{align}
where $P_M$ is the UAV transmit power and $B_M$ is the inter-UAV bandwidth.

\subsection{Computing Model}
In each time slot $t$, MT $k$ generates a computing task characterized by $\{J_k(t), C_k\}$, where $J_k(t)$ denotes the input data size (in bits) and $C_k$ denotes the required CPU cycles per bit. This slot-level task-generation model is used to describe the time-varying computing demands of MTs in a tractable manner. Since the task is divisible, MT $k$ can offload a portion of the task to a UAV. 
Let $\gamma_k(t)\in[0,1]$ denote the offloading ratio, where $\gamma_k(t)$ is the fraction offloaded and $1-\gamma_k(t)$ is computed locally.
Accordingly, the local computation time of MT $k$ in slot $t$ is
\begin{align}
D_k^{{\rm local}}(t)=\frac{(1-\gamma_k(t))J_k(t)C_k}{f_k^{\rm loc}},
\end{align}
where $f_k^{\rm loc}$ is the computing power of MT $k$ (CPU cycles/s).
\par

\vspace{-3pt}
Let $z_{k,m}(t)\in\{0,1\}$ indicate the association decisions, with $\sum_{m\in\mathcal{M}(t)} z_{k,m}(t)=1$ since each MT can access only one UAV per slot. 
For notational simplicity, for each MT $k$ in slot $t$, we use $m_0$ to denote its unique associated UAV, i.e., $z_{k,m_0}(t)=1$.
The uplink transmission time for MT $k$ to offload task data to UAV $m$ in slot $t$ is
\begin{align}
D_{k,m}^{{\rm comm}}(t)=\frac{\gamma_k(t)\,z_{k,m}(t)\,J_k(t)}{r_{k,m}(t)}.
\end{align}
\par

\vspace{-7pt}
Due to the uneven spatial distribution of MTs, the number of MTs associated with each UAV may vary, which causes load imbalance. To alleviate this, a UAV can migrate a portion of its offloaded workload to nearby UAVs with available computing resources.
Specifically, for the task offloaded by MT $k$ to UAV $m_0$ in slot $t$, UAV $m_0$ may forward a fraction of this workload to UAV $m$. 
Let $\rho_{k,m_0,m}(t)\in[0,1]$ denote the migration ratio from UAV $m_0$ to UAV $m$, satisfying $\sum_{m\in\mathcal{M}(t)} \rho_{k,m_0,m}(t)=1, \forall k\in\mathcal{K},~ t\in\mathcal{T}$.
Accordingly, the transmission time for forwarding the migrated task data from UAV $m_0$ to UAV~$m$~is
\begin{align}
D_{k,m_0,m}^{{\rm tran}}(t)=\frac{J_k(t)\,\gamma_k(t)\,\rho_{k,m_0,m}(t)}{r_{m_0,m}(t)}.
\end{align}
To unify the notation, we define $D_{k,m_0,m}^{\rm tran}(t)=0$ when $m=m_0$, corresponding to the case where the offloaded workload fraction is scheduled for execution at its associated UAV $m_0$ and therefore no inter-UAV forwarding is required.
\par

In time slot $t$, the computation time at UAV $m$ for processing the workload of MT is
\begin{align}
D_{k,m}^{\rm comp}(t)
= \frac{J_k(t)C_k\,\gamma_k(t)\,\rho_{k,m_0,m}(t)}{f_{k,m}(t)},\label{eq:d_comp}
\end{align}
where $f_{k,m}(t)$ denotes the computing power (CPU cycles/s) allocated by UAV $m$ to process MT $k$'s workload in slot $t$ that satisfies $\sum\limits_{k \in {\mathcal{K}}} {{f_{k,m}}\left( t \right)}  \le f_m^{\max}$, in which $f_m^{\max }$ is the maximum computing power of the UAV $m$.
Here, the inner summation over $\tilde m$ only serves to select the unique associated UAV of MT $k$ through the one-hot association variable $z_{k,\tilde m}(t)$.
Due to the relatively small size of computation results, we ignore the downlink transmission, including inter-UAV result delivery. 
\par

Therefore, for MT $k$, the processing time of the task generated in slot $t$ is
\begin{align}
\small
&D_k(t)=\max\!\Bigg(
D_k^{\rm local}(t),~
\Big[D_{k,m_0}^{\rm comm}(t)\notag\\
&\qquad\qquad\qquad
+\max_{m\in\mathcal{M}(t)}\big( D_{k,m_0,m}^{\rm tran}(t)+D_{k,m}^{\rm comp}(t)\big)
\Big]
\Bigg).
\end{align}
In the above expression, local computing and uplink offloading are performed in parallel at MTs, while the offloaded workload may be split and processed in parallel across multiple UAVs. Hence, the completion time of the offloaded part is determined by the slowest execution branch.
Moreover, the duration of slot $t$ can be set according to the maximum task processing time among all MTs, i.e., $\tau_t=\min\left\{\delta,\, \max_{k\in\mathcal{K}} D_k(t)\right\}$. Therefore, the slot duration in slot $t$ is determined by the corresponding task processing requirement while remaining upper bounded by $\delta$, and the considered setting does not involve unfinished tasks carried over to subsequent slots.
\par

For UAV $m$, the total amount of task data executed in time slot $t$ is
\begin{align}
W_m(t)
= \sum_{k\in\mathcal{K}} J_k(t)\,\gamma_k(t)\,\rho_{k,m_0,m}(t).
\end{align}
That is, in each summation term, $m_0$ denotes the associated UAV of the corresponding MT $k$, and $W_m(t)$ accounts for all workload fractions finally executed by UAV $m$.

\vspace{-18pt}
\subsection{Energy Consumption Model}
In time slot $t$, the communication energy consumption for MT $k$ to offload task data to UAV $m$ is
\begin{equation}
    E_{k,m}^{{\rm comm}}(t)=D_{k,m}^{{\rm comm}}(t)P_K,
\end{equation}
while the local computation energy consumption of MT $k$ is expressed as
\begin{align}
E_k^{{\rm local}}(t)=\varphi_k \big(f_k^{\rm loc}\big)^3 D_k^{{\rm local}}(t),
\end{align}
where $\varphi_k$ is the effective capacitance coefficient of MT $k$.
Therefore, the total energy consumption of MT $k$ in slot $t$~is
\begin{align}
E_k(t)=E_k^{{\rm local}}(t)+\sum_{m\in\mathcal{M}(t)} E_{k,m}^{{\rm comm}}(t).
\end{align}
\par

According to the rotary-wing UAV power model, the propulsion power of UAV $m$ in time slot $t$ is given by
\begin{equation}
    \begin{aligned}
        p_m(t)&=U_1\left(1+\frac{3v_m^2(t)}{v_{tip}^2}\right) \\
        &+U_2\sqrt{\sqrt{U_3+\frac{v_m^4(t)}{4}}-\frac{v_m^2(t)}{2}}
        +U_4 v_m^3(t)+U_5,
    \end{aligned}
\end{equation}
where $U_1, U_2, U_3,U_4,U_5$ are constants related to aerodynamics and UAV hardware, and $v_{tip}$ is the rotor tip speed.
Thus, the propulsion energy consumption of UAV $m$ in slot $t$ is represented as 
\begin{equation}
    E_m^{{\rm fly}}(t)=p_m(t)\tau_t.
\end{equation}
In time slot $t$, the communication energy consumption for forwarding the migrated workload from UAV $m_0$ to UAV $m$~is 
\begin{equation}
    E_{k,m_0,m}^{{\rm tran}}(t)=D_{k,m_0,m}^{{\rm tran}}(t)P_M.
\end{equation}
The computation energy consumption at UAV $m$ for processing MT $k$'s workload is given by
\begin{align}
E_{k,m}^{{\rm comp}}(t)=\varphi_m \big(f_{k,m}(t)\big)^3 D_{k,m}^{{\rm comp}}(t),
\end{align}
where $\varphi_m$ is the effective capacitance coefficient of UAV $m$.
\par

Therefore, the total energy consumption of UAV~$m$ in slot~$t$ is expressed as
\begin{align}
E_m(t)=E_m^{{\rm fly}}(t)+\sum_{k\in\mathcal{K}}\Big(&\sum_{m'\in\mathcal{M}(t)} z_{k,m}(t)\,E_{k,m,m'}^{{\rm tran}}(t) \notag\\
&+E_{k,m}^{{\rm comp}}(t)\Big).
\end{align}
Here, $z_{k,m}(t)=1$ indicates that the currently considered UAV $m$ is the associated UAV of MT $k$ in slot $t$, and thus incurs the corresponding workload-forwarding energy.
In the above expression, the first term inside the summation captures the transmission energy for forwarding workload fractions from UAV $m$ to UAV $m'$, where $m'$ denotes the execution UAV, while the second term inside the summation captures the computation energy consumed by UAV $m$ when it executes the assigned workload.

\subsection{Stackelberg Game Formulation}
We consider a distributed UAV-enabled MEC system where UAVs provide computing services to MTs. 
In each slot, the UAV and MT positions are first updated, and the corresponding channel conditions are determined. Then, the UAVs make the slot-level service decisions and announce their unit prices. Based on the current system state and the announced decisions, each MT determines its offloading ratio and requested computing resources. After that, the offloaded workload is transmitted to the associated UAV and, if necessary, further forwarded to other UAVs for collaborative execution. Finally, the task delay, energy consumption, UAV utility, and MT overhead are updated for the current slot.
Such a slot-level interaction naturally motivates a Stackelberg game formulation, where UAVs act as leaders and MTs act as followers.
At the leader level, the UAVs jointly determine the trajectories, task migration ratios, MT-UAV association decisions, and unit prices for computing service. Given these leader-level decisions, each MT determines its offloading ratio and requested computing resources at the follower level.
UAVs aim to maximize benefits from resource provisioning, while MTs seek to acquire sufficient computing resources at minimal overhead. 
In general, higher prices discourage offloading and lead MTs to execute more workload locally, while lower prices incentivize more aggressive offloading. Let $b_m(t)$ denote the unit price announced by UAV $m$ in slot $t$.
For each UAV, the instantaneous benefit in slot $t$ balances three factors: the utility gained from the workload finally executed by itself, the corresponding energy cost, and the revenue obtained by providing computing resources to MTs.
These decisions are constrained by practical limitations such as onboard energy and computing capacity.
Accordingly, the benefit of UAV $m$ in time slot $t$ is~given~by
\begin{align}
U_m(t)=\omega^{\rm W} W_m(t)-\omega^{\rm E} E_m(t)+b_m(t)\sum_{k\in\mathcal{K}} f_{k,m}(t),
\end{align}
where $\omega^{\rm W}$ and $\omega^{\rm E}$ are the weights associated with the executed task data and the energy consumption, respectively, which are used to balance the incentive of executing more workload against the corresponding energy cost in the UAV-side objective.
$W_m(t)$ and $E_m(t)$ denote the workload finally executed by UAV $m$ and its overall energy consumption.
\par

Therefore, UAV $m$ maximizes its long-term benefits by jointly optimizing its trajectory, task migration ratios, UAV access selection, and the unit price of computing resources. The leader-level optimization problem is formulated as
\begin{subequations}\label{prob:P1}
\begin{align}
{\rm(P1):}\quad &\max_{\{\Lambda_m\}}~ \sum_{t\in\mathcal{T}} U_m(t),\\
\text{s.t.}\quad
& 0 \le \rho_{k,m,m'}(t) \le 1,
&& \forall k,\forall m,\forall m',\forall t, \label{P1_rho_bd}\\
& \sum_{m'\in\mathcal{M}(t)} \rho_{k,m,m'}(t)=1,
&& \forall k,\forall m,\forall t, \label{P1_rho_sum}\\
& z_{k,m}(t)\in\{0,1\},
&& \forall k,\forall m,\forall t, \label{P1_z_bin}\\
& \sum_{m\in\mathcal{M}(t)} z_{k,m}(t)=1,
&& \forall k,\forall t, \label{P1_z_sum}\\
& q_m(1)=q_m^{\rm ini},
&& \forall m, \label{P1_init}\\
& 0\le v_m(t)\le V^{\max},
&& \forall m,\forall t, \label{P1_v_bd}\\
& 0\le \alpha_m(t)<2\pi,
&& \forall m,\forall t, \label{P1_a_bd}\\
& d_{m,m'}(t)\ge d^{\min},
&& \forall m\ne m',\forall t, \label{P1_safe}\\
& \sum_{t\in\mathcal{T}} E_m(t)\le E_m^{\max},
&& \forall m, \label{P1_energy}\\
& b_m(t)\ge 0,
&& \forall m,\forall t, \label{P1_price}
\end{align}
\end{subequations}
where $\Lambda_m=\big(v_m(t),\alpha_m(t),\rho_{k,m,m'}(t),z_{k,m}(t),b_m(t)\big)$ denotes the leader-level decision variables of UAV $m$\footnote{The notation $\rho_{k,m,m'}(t)$ is kept in its general form in the optimization problem, where $m$ and $m'$ denote the source UAV and the execution UAV, respectively. This is because the leader-level formulation describes the migration decision from the UAV-level perspective. By contrast, the notation $m_0$ introduced earlier is only used as a local notation for the unique associated UAV of MT $k$ in slot $t$ when expressing the resulting workload, delay, and energy terms from the task-level perspective.}.
Constraints \eqref{P1_rho_bd}--\eqref{P1_z_sum} guarantee feasible migration and association decisions. Constraints \eqref{P1_init}--\eqref{P1_safe} capture trajectory-related feasibility, including the initial position, mobility bounds, and safety distance. Moreover, \eqref{P1_energy} and \eqref{P1_price} impose the UAV energy budget and the non-negative pricing constraint, respectively.
\par

For MT $k$, we define the MT overhead as a sum of the task processing delay and the payment for purchased computing resources. By jointly optimizing the offloading ratio and the requested computing resources, the MT overhead minimization problem is formulated as
\begin{subequations}\label{prob:P2}
\begin{align}
{\rm(P2):}\quad
&\min_{\{\Gamma_k\}}~\sum_{t\in\mathcal{T}}\Big( D_k(t) + \sum_{m\in\mathcal{M}(t)}b_m(t)~f_{k,m}(t) \Big),\\
\text{s.t.}\quad
& 0 \le f_{k,m}(t) \le z_{k,m}(t)\,f_m^{\max},\qquad \forall m,\forall t, \label{P2_f_bd}\\
& \sum_{k\in\mathcal{K}} f_{k,m}(t)\le f_m^{\max}, \qquad \quad \qquad \forall m,\forall t, \label{P2_cap}\\
& 0\le \gamma_k(t)\le 1,\qquad \qquad \quad \qquad \forall t, \label{P2_gamma}\\
& E_k(t)\le E_k^{\max},\qquad \quad \qquad \qquad \forall t, \label{P2_E}
\end{align}
\end{subequations}
where ${\Gamma_k}=\big(f_{k,m}(t),\gamma_k(t)\big)$ denotes the decision variables of MT $k$.
Constraint \eqref{P2_f_bd} enforces the feasibility of the requested computing resources.
\eqref{P2_cap} captures the computing capacity limit of each UAV.
\eqref{P2_gamma} bounds the offloading ratio within $[0,1]$.
\eqref{P2_E} ensures that the energy consumption of MT $k$ in each slot does not exceed its energy budget.

\section{Distributed Hierarchical Optimization Algorithm}\label{sec:alg}

To achieve efficient MEC computing resource allocation, meeting the demands of latency-sensitive tasks, we propose the decision-making algorithms for the leader and the followers of the formulated Stackelberg game optimization model based on the MADRL method and the Lagrange multiplier method, respectively.

\subsection{Leader Decision Algorithm Based on MAPPO}
Problem $(\rm{P1})$ involves coupled trajectory-related decisions across time slots and strong interactions among UAVs through MT association and inter-UAV task migration. 
Moreover, MT mobility and stochastic task arrivals lead to a time-varying environment, and latency-sensitive services require online decision-making. Under such conditions, traditional optimization methods often fail to provide efficient solutions. 
Therefore, we model the UAV decision-making process as a Markov decision process and adopt MAPPO, whose centralized-training-and-decentralized-execution paradigm is well suited to problem $(\rm{P1})$ with strongly coupled multi-UAV decisions in a time-varying environment, while still supporting distributed execution based on local observations after training.

\par
We transform $(\rm{P1})$ into a partially observable Markov decision process (POMDP) with $M(t)$ agents.
The POMDP is defined by the state space ${\cal S}$, observation spaces ${\cal O}=\{{\cal O}_1,\ldots,{\cal O}_{M(t)}\}$, action spaces ${\cal A}=\{{\cal A}_1,\ldots,{\cal A}_{M(t)}\}$, and reward functions ${\cal R}=\{{\cal R}_1,\ldots,{\cal R}_{M(t)}\}$. The state ${\cal S}$ captures global environment information, including UAV locations, MT locations, and task-related information, while each UAV $m$ observes a local observation ${o}_m(t)\in{\cal O}_m$ and takes an action ${a}_m(t)\in{\cal A}_m$ according to its policy $\pi_m$. The specific definitions of state, observation, action, and reward are given as follows:

1) \textit{State space}: In time slot $t$, the global information of the environment can be represented as
$s(t)=\big(\mathbf{q}(t),\mathbf{w}(t),\mathbf{J}(t)\big),$ where $\mathbf{q}(t)=\{(x_m(t),y_m(t))\}_{m\in\mathcal{M}(t)}$ denotes the UAV locations, $\mathbf{w}(t)=\{(x_k(t),y_k(t))\}_{k\in\mathcal{K}}$ denotes the MT locations, and
$\mathbf{J}(t)=\{J_k(t)\}_{k\in\mathcal{K}}$ denotes the task input sizes.

2) \textit{Observation space}: In time slot $t$, the local information observed by UAV $m$ can be expressed as $o_m(t)=\big(q_m(t),\mathbf{w}(t),\mathbf{J}(t)\big),$ which includes the location of UAV $m$, the locations of all MTs, and the task input sizes. In the considered setting, the MT-side information is available to the UAVs on a slot basis, while the positions of the other UAVs are not included in the local observation.

3) \textit{Action space}: In time slot $t$, the action of UAV $m$ can be represented as
$a_m(t)=\big(v_m(t),\alpha_m(t),\boldsymbol{\rho}_m(t),\mathbf{z}_m(t),b_m(t)\big),$ where $\boldsymbol{\rho}_m(t)=\{\rho_{k,m,m'}(t)\}_{k\in\mathcal{K},m'\in\mathcal{M}(t)}$ denotes migration ratios from UAV $m$ and $\mathbf{z}_m(t)=\{z_{k,m}(t)\}_{k\in\mathcal{K}}$ denotes association decisions. 
To ensure action feasibility, the continuous outputs are mapped onto their admissible ranges, and $\boldsymbol{\rho}_m(t)$ is further normalized to satisfy \eqref{P1_rho_sum}.

4) \textit{Reward function}: In time slot $t$, the reward function of UAV $m$ can be expressed as ${r_m}\left( t \right) = {\omega ^{\rm{W}}}{W_m}\left( t \right) - {\omega ^{\rm{E}}}{E_m}\left( t \right) + {b_m}\left( t \right)\sum\limits_{k \in {\cal K}} {{f_{k,m}}\left( t \right)} $, denotes the total benefit obtained by UAV $m$ in time slot $t$.
\par

To fit the considered distributed multi-UAV MEC setting, the MAPPO formulation is specified through the above state, observation, action, and reward design, together with the corresponding feasibility handling for the leader-level decisions.
In MAPPO, each agent is composed of two actor networks (${\pi _{{\theta _m}}}$ and ${\pi _{\theta _m^{{\rm{old}}}}}$), a critic network ${Q_{\theta _m^Q}}$, and an experience buffer. 
During the training phase, a central controller collects and evaluates global information about the environment to achieve centralized training. During the execution phase, each agent implements distributed execution based on its locally deployed actor network, without requiring a centralized controller for real-time decision making.
Specifically, the intelligent agent inputs the observed result ${o_m}\left( t \right)$ into the actor network, and then obtains the action ${a_m}\left( t \right)$ and reward ${r_m}\left( t \right)$ of the current training batch, and stores the experience in the experience buffer. 
At the end of each training batch, the agent randomly extracts a certain batch of data from the buffer and calculates the advantage function value ${\hat A_{{\theta _m}}}\left( {o\left( t \right)} \right) = \sum\nolimits_{l = 0}^{T - t} {{\gamma ^l}} \eta \left( {t + l} \right)$, where $\gamma $ is the discount factor and $\eta \left( {t + l} \right)$ represents the temporal difference error, defined as
\begin{align}
\eta \left( t \right) = R\left( t \right) + \gamma {Q_{\theta _m^Q}}\left( {o\left( {t + 1} \right)} \right) - {Q_{\theta _m^Q}}\left( {o\left( t \right)} \right).
\end{align}
Among them, ${Q_{\theta _m^Q}}\left( {o\left( t \right)} \right)$ represents the output of the critic network ${Q_{\theta _m^Q}}$ when its input is $o\left( t \right)$.

The loss function of the actor network is
\begin{align}
J_{\theta_m}
&=\mathbb{E}\Big[ \min \Big(
u_{\theta_m}(t)\,\hat A_{\theta_m}(t), \notag\\
&\qquad \mathrm{clip}\!\big(u_{\theta_m}(t),1-\varepsilon,1+\varepsilon\big)\,\hat A_{\theta_m}(t)
\Big) \Big].
\end{align}
Among them, ${u_{{\theta _m}}}\left( t \right)$ represents the ratio of the probability of the actor network strategy corresponding to agent $m$:
\begin{align}
{u_{{\theta _m}}}\left( t \right) = \frac{{{\pi _{{\theta _m}}}\left( {a_m\left( t \right)\big|o_m\left( t \right)} \right)}}{{{\pi _{\theta _m^{{\rm{old}}}}}\left( {a_m\left( t \right)\big|o_m\left( t \right)} \right)}}.
\end{align}
where ${\pi _{{\theta _m}}}\left( {a\left( t \right)|o\left( t \right)} \right)$ represents the probability that the actor network ${\pi _{{\theta _m}}}$ outputs $a\left( t \right)$ when the input is $o\left( t \right)$. Similarly, the same applies to ${\pi _{\theta _m^{{\rm{old}}}}}\left( {a\left( t \right)|o\left( t \right)} \right)$.

To avoid excessive parameter updates in the actor network, a clip function is defined to constrain ${u_{{\theta _m}}}\left( t \right)$ between $1 - \varepsilon $ and $1 + \varepsilon $:
{\small
\begin{align}
{\rm{clip}}\left( {{u_{{\theta _m}}}\left( t \right),1 - \varepsilon ,1 + \varepsilon } \right) = \left\{ \begin{array}{l}
1 - \varepsilon ,{\rm{     if  }}{u_{{\theta _m}}}\left( t \right) \le 1 - \varepsilon ,\\
1 + \varepsilon ,{\rm{     if  }}{u_{{\theta _m}}}\left( t \right) \ge 1 + \varepsilon ,\\
{u_{{\theta _m}}}\left( t \right),{\rm{   otherwise}}{\rm{.}}
\end{array} \right.
\end{align}}

Every certain training batch, training data is retrieved from the experience buffer, and then centralized updates of all actor networks and critic networks are completed in the central controller. The cumulative discount reward is defined as $G\left( t \right) = \sum\nolimits_{l = 0}^{T - t} {{\gamma ^l}R\left( {t + l} \right)} $, and the loss function of the critic network is defined as
\begin{align}
{J_{\theta _m^Q}} = \mathbb{E}\!\left[ {{{\left( {{Q_{\theta _m^Q}}\left( {s\left( t \right)} \right) - G\left( t \right)} \right)}^2}} \right].
\end{align}

\subsection{Follower decision algorithm based on two-stage iteration}
For problem $(\rm{P2})$, the per-slot decisions are separable across time slots. Hence, we solve the single-slot problem and omit the slot index $t$ for brevity. 
Under the given leader-side association, migration, and pricing decisions, the follower-side problem reduces to the joint optimization of each MT's offloading ratio and requested computing resources. Based on this structure, we develop a two-stage iterative procedure with updates derived via Lagrange duality and KKT conditions.
To linearize the max-type delay expression, we introduce the auxiliary variables $\{\chi_{k,m}\}_{m\in\mathcal{M}}$ and $G_k$ such that
\begin{align}
 &\chi_{k,m}\ge D_{k,m}^{\rm comp}, \qquad &&\forall m\in\mathcal{M}, \label{eq:chi1}\\
 &\chi_{k,m}\ge D_{k,m,m'}^{\rm tran} + D_{k,m'}^{\rm comp}, \qquad &&\forall m,m'\in\mathcal{M}, \label{eq:chi2}\\
 &G_k\ge D_k^{\rm local}, \label{eq:G1}\\
 &G_k\ge D_{k,m}^{\rm comm} + \chi_{k,m}, \qquad &&\forall m\in\mathcal{M}. \label{eq:G2}
\end{align}
Then the follower subproblem for MT $k$ can be reformulated as
\begin{subequations}\label{prob:P3}
\begin{align}
\rm{(P3):}\quad &\min_{\gamma_k,\{f_{k,m}\},G_k,\{\chi_{k,m}\}}
\quad G_k + \sum_{m\in\mathcal{M}} b_m f_{k,m}, \label{prob:P3-obj}\\
\text{s.t.}\quad
&\quad \eqref{eq:chi1}\text{--}\eqref{eq:G2}, \label{prob:P3-epi}\\
&\quad f_{k,m}\ge 0,\quad \sum_{k=1}^{K} f_{k,m} \le f_m^{\max},\quad \forall m\in\mathcal{M}, \label{prob:P3-cap}\\
&\quad 0\le \gamma_k \le 1,\quad \tau_t\le \delta,\quad E_k(\gamma_k)\le E_k^{\max}. \label{prob:P3-box}
\end{align}
\end{subequations}
In \eqref{prob:P3}, the actual decision variables are $(\gamma_k,\{f_{k,m}\})$, while $G_k$ and $\{\chi_{k,m}\}$ are auxiliary slack variables introduced by the epigraph reformulation.

\begin{remark}
Problem $(\rm{P3})$ admits a non-cooperative game interpretation among MTs, where each MT $k$
selects $u_k=(\gamma_k,\{f_{k,m}\}_{m\in\mathcal{M}})$ to minimize its individual overhead
$G_k' \triangleq G_k + \sum_{m\in\mathcal{M}} b_m f_{k,m}$.
A strategy profile ${\bf u}^*=\{u_k^*\}_{k\in\mathcal{K}}$ is a Nash equilibrium if
$G_k'(u_k^*,u_{-k}^*) \le G_k'(u_k,u_{-k}^*)$ for all feasible $u_k$ and $k\in\mathcal{K}$ \cite{9732351}.
\end{remark}
\par

Accordingly, we employ distributed best-response updates, where each MT alternates between
\eqref{eq:gamma-star} and \eqref{eq:f-star} while treating other MTs' decisions as fixed.
To obtain a low-complexity solver, we adopt a two-stage alternating procedure to solve \eqref{prob:P3} for each MT $k$.

\emph{(1) Update $\gamma_k$ with fixed $\{f_{k,m}\}$:}
With fixed computing resources $\{f_{k,m}\}$, problem \eqref{prob:P3} reduces to a convex optimization with a single scalar decision $\gamma_k$ and linear epigraph constraints. The Lagrangian function can be written as
\begin{align}
\mathcal{L}^{(\gamma)}
&= G_k
+ \zeta\!\left(D_k^{\rm local}-G_k\right) \notag\\
&+ \sum_{m\in\mathcal{M}}\upsilon_m\!\left(D_{k,m}^{\rm comm}+\chi_{k,m}-G_k\right) \notag\\
&\quad + \sum_{m\in\mathcal{M}}\sum_{m'\in\mathcal{M}}
\vartheta_{m,m'}\!\left(D_{k,m,m'}^{\rm tran}+D_{k,m'}^{\rm comp}-\chi_{k,m}\right) \notag\\
&\quad - \underline{\theta}_k\,\gamma_k + \bar{\theta}_k(\gamma_k-1)
+ \varepsilon\!\left(E_k-E_k^{\max}\right),
\label{eq:L-P3-1}
\end{align}
where $\zeta\ge 0$, $\{\upsilon_m\ge 0\}$, $\{\vartheta_{m,m'}\ge 0\}$, $\underline{\theta}_k\ge 0$, $\bar{\theta}_k\ge 0$, and $\varepsilon\ge 0$ are dual variables. By applying the KKT conditions, the stationarity with respect to $\gamma_k$ yields a scalar equation, which leads to a closed-form update
\begin{align}
\gamma_k^* = \min\!\left\{\gamma_k^{\rm E},\ \frac{C_k}{\gamma_k^{\rm den}}\right\}.
\label{eq:gamma-star}
\end{align}
Here $\gamma_k^{\rm E}$ is the maximum feasible offloading ratio induced by the energy constraint. Under the commonly used energy model where the local CPU energy scales with $(f_k^{\rm loc})^2$ and the uplink energy is proportional to the transmission time, we can obtain
\begin{align}
\gamma_k^{\rm E}
=\min\!\left\{
\frac{E_k^{\max}- J_k C_k \varphi_k (f_k^{\rm loc})^2}
{\sum_{m\in\mathcal{M}} z_{k,m}\frac{J_k P_k}{r_{k,m}}- J_k C_k \varphi_k (f_k^{\rm loc})^2},
\ 1
\right\},
\label{eq:gamma-E}
\end{align}
and $\gamma_k^{\rm den}$ collects the coefficients of $\gamma_k$ in the end-to-end delay expression as
\begin{align}
\gamma_k^{\rm den}
&= C_k
+ f_k^{\rm loc}\sum_{m\in\mathcal{M}} \frac{z_{k,m}}{r_{k,m}} \notag\\
&+ f_k^{\rm loc}\sum_{m\in\mathcal{M}}\sum_{m'\in\mathcal{M}}
\left(
\frac{z_{k,m}\rho_{k,m,m'}}{r_{m,m'}}
+ \frac{C_k z_{k,m'}\rho_{k,m',m}}{f_{k,m'}}
\right).
\label{eq:gamma-den}
\end{align}
Specifically, we substitute the explicit delay and energy models into $\partial \mathcal{L}^{(\gamma)}/\partial \gamma_k=0$,
and then collect the coefficients of $\gamma_k$ to obtain $\gamma_k^{\rm den}$, while $\gamma_k^{\rm E}$ is derived from the energy feasibility $E_k\le E_k^{\max}$.
 
\emph{(2) Update $\{f_{k,m}\}$ with fixed $\gamma_k$:}
With fixed $\gamma_k$, problem \eqref{prob:P3} is convex in $\{f_{k,m}\}$ under the per-UAV capacity constraints.
The dependence on $f_{k,m}$ comes from the linear price term $b_m f_{k,m}$ and the computation delay in~\eqref{eq:d_comp}.
Let $\lambda_m\ge 0$ be the dual variable associated with the capacity constraint $\sum_{k=1}^{K} f_{k,m}\le f_m^{\max}$.
Moreover, $D_{k,m}^{\rm comp}$ enters the epigraph constraints via $\chi_{k,\bar m}\ge D_{k,\bar m,m}^{\rm tran}+D_{k,m}^{\rm comp}$.
Focusing on the terms involving $\{f_{k,m}\}_{k\in\mathcal{K}}$ for UAV $m$, the Lagrangian function yields
\begin{align}
\mathcal{L}^{(f)}_m
= \sum_{k\in\mathcal{K}} (b_m+\lambda_m)\, f_{k,m}
+ \sum_{k\in\mathcal{K}} \omega_{k,m}\, D_{k,m}^{\rm comp},
\label{eq:L-fm}
\end{align}
Under \eqref{eq:chi1}--\eqref{eq:chi2}, the aggregated weight is
$\omega_{k,m} = \phi_m + \sum_{\bar m\in\mathcal{M}} \vartheta_{\bar m,m}$,
where $\phi_m\ge 0$ and $\vartheta_{\bar m,m}\ge 0$ are the multipliers of \eqref{eq:chi1} and \eqref{eq:chi2}, respectively.

Applying the stationarity condition $\partial \mathcal{L}^{(f)}_m/\partial f_{k,m}=0$ gives $(b_m+\lambda_m) - \omega_{k,m}\frac{J_k C_k \sum_{m'\in\mathcal{M}} \big( z_{k,m'}\gamma_k\rho_{k,m',m} \big)}{f_{k,m}^2}=0$,
which yields
\begin{align}
f_{k,m}^*
= \sqrt{\frac{\omega_{k,m}\,J_k C_k \sum\limits_{m'\in\mathcal{M}} (z_{k,m'}\gamma_k\rho_{k,m',m})}{b_m+\lambda_m}},
\label{eq:f-star}
\end{align}
and it satisfies $f_{k,m}^*\ge 0$ since $b_m+\lambda_m>0$ and the numerator is nonnegative.
For each UAV $m$, $\lambda_m$ is determined by the complementary slackness condition with
$\sum_{k=1}^{K} f_{k,m}^*\le f_m^{\max}$.

\par 
\emph{(3) Two-stage iteration:}
For each MT $k$, we alternately update $\gamma_k$ and $\{f_{k,m}\}$ according to \eqref{eq:gamma-star} and \eqref{eq:f-star} until the local change is below a tolerance.
MTs then perform best-response updates in an outer loop until the strategy profile stabilizes.
The procedure can be implemented in a distributed manner since each MT only exchanges the current decisions with other MTs and no centralized controller is required.
The overall procedure is summarized in Algorithm~\ref{A1}.

\begin{algorithm}[t]
\small
\caption{Follower decision algorithm based on two-stage iteration}
\label{A1}
\begin{algorithmic}[1]
\STATE Set tolerance $\epsilon$, outer iteration index $j=0$, and maximum outer iterations $J^{\max}$.
\STATE Initialize $\gamma_k^{(0)}$ and $f_{k,m}^{(0)}$ for all $k\in\mathcal{K}$ and $m\in\mathcal{M}$.
\REPEAT
  \FOR{$k=1$ to $K$}
    \STATE Set inner index $i=0$; set $\gamma_k^{i}\leftarrow \gamma_k^{(j)}$, $f_{k,m}^{i}\leftarrow f_{k,m}^{(j)}$, $\forall m$.
    \REPEAT
      \STATE Update $\gamma_k^{i+1}$ by \eqref{eq:gamma-star} with fixed $\{f_{k,m}^{i}\}$.
      \STATE Update $\{f_{k,m}^{i+1}\}$ by \eqref{eq:f-star} with fixed $\gamma_k^{i+1}$, where $\{\lambda_m\}$ are obtained to satisfy $\sum_{k=1}^K f_{k,m}^{i+1}\le f_m^{\max}$ (e.g., by bisection).
      \STATE $i \leftarrow i+1$.
    \UNTIL $\left|\gamma_k^{i}-\gamma_k^{i-1}\right| + \sum_{m\in\mathcal{M}}\left|f_{k,m}^{i}-f_{k,m}^{i-1}\right| \le \epsilon$
    \STATE Set $\gamma_k^{(j+1)}\leftarrow \gamma_k^{i}$ and $f_{k,m}^{(j+1)}\leftarrow f_{k,m}^{i}$, $\forall m$.
  \ENDFOR
  \STATE $j \leftarrow j+1$.
\UNTIL $\max\limits_{k\in\mathcal{K}}\!\left(\left|\gamma_k^{(j)}-\gamma_k^{(j-1)}\right| + \sum_{m\in\mathcal{M}}\left|f_{k,m}^{(j)}-f_{k,m}^{(j-1)}\right|\right)\le \epsilon$ or $j \ge J^{\max}$
\end{algorithmic}
\end{algorithm}

\subsection{Distributed Stackelberg Game Algorithm Based on MADRL}
By integrating the MAPPO-based leader optimization with the two-stage follower solver, we obtain a distributed Stackelberg game algorithm based on MADRL under a centralized-training-and-distributed-execution (CTDE) paradigm.

\begin{figure}[!t]
\centering
\includegraphics[width= \columnwidth]{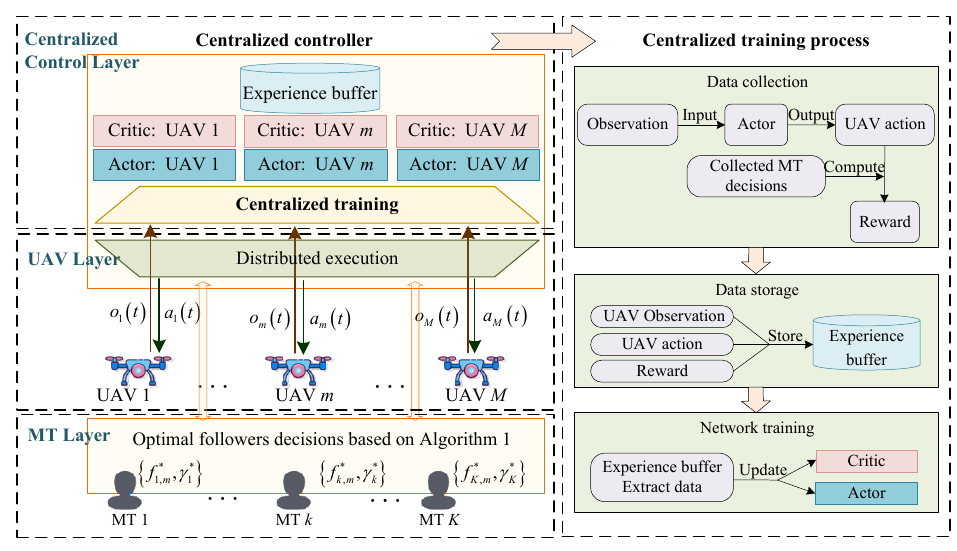}
\caption{Training diagram of distributed Stackelberg game algorithm based on MADRL.}
\label{train}
\end{figure}

\emph{Centralized training:}
At each time slot, each UAV collects its local observation and uploads it to a centralized trainer, which constructs the joint state for critic evaluation.
The UAV actions are sampled from the actor networks, and the follower decisions (MT best responses) are computed via Algorithm~\ref{A1}.
The environment then returns the reward and the next state, and the resulting transitions are stored in an experience buffer.
When a training batch is ready or the last slot of a period is reached, the trainer updates the actor and critic networks of all UAV agents.
This centralized training stage incurs communication overhead between the UAVs and the trainer due to observation uploading and updated parameter feedback. Since such communication only appears periodically during training and does not affect the runtime distributed execution procedure, it is regarded as a training-side implementation cost in this work.
The training flow is illustrated in Fig.~\ref{train}, and the detailed procedure is given in Algorithm~\ref{A2}.

\begin{figure}[!t]
\centering
\includegraphics[width= \columnwidth]{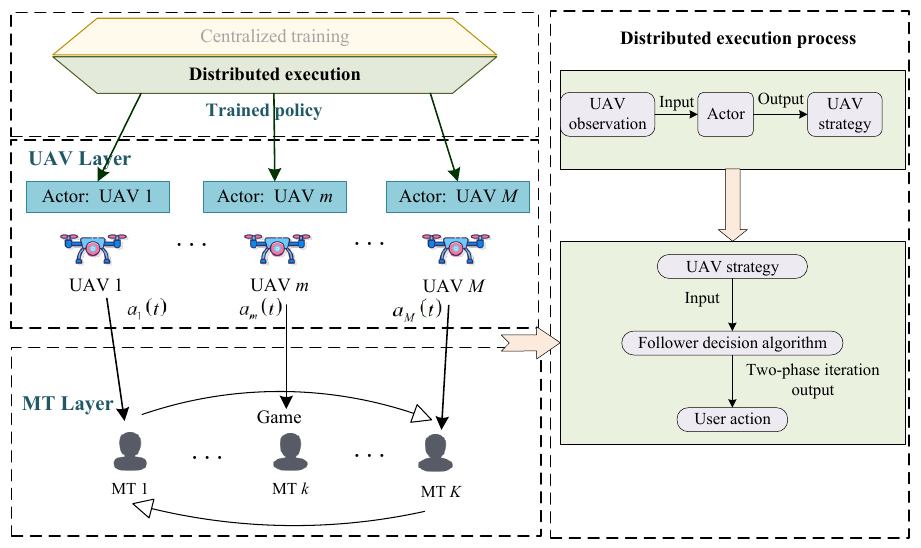}
\caption{Execution diagram of distributed Stackelberg game algorithm based on MADRL}
\label{execute}
\end{figure}

\emph{Distributed execution:}
Each UAV runs its trained actor network locally and selects actions based on its own observation.
At runtime, each UAV sends its slot-level service information to the related MTs through lightweight control messages,which may include the unit price $b_m(t)$ and the association outcome $z_{k,m}(t)$.
Based on the received service information and scalar capacity feedback from the involved UAVs, the MTs perform the follower-side updates in Algorithm~\ref{A1} to obtain their offloading ratios $\{\gamma_k(t)\}$ and requested computing resources $\{f_{k,m}(t)\}$. The resulting follower-side decisions are then sent to the associated UAVs.
When collaborative execution is required, the associated UAV further sends coordination messages to the selected execution UAVs, which may include the task identifier, the migration ratio $\rho_{k,m,m'}(t)$, and the workload fraction to be forwarded.
The additional communication overhead introduced by Algorithm~\ref{A1} is limited to scalar-level control exchange during the iterative decision update. In particular, the bisection-based determination of $\{\lambda_m\}$ only requires capacity-related scalar feedback from the involved UAVs. Hence, this overhead mainly scales with the number of involved MTs, the number of available UAVs, and the iteration budgets, while no raw task payloads, global observations, or neural network parameters are exchanged in this iterative decision update.
No centralized controller is involved in this stage, and no global observations are collected for decision-making. Therefore, the runtime process only relies on local observations and slot-level control signaling among the involved UAVs and MTs over the available UAV-MT and inter-UAV links.
The execution process is illustrated in Fig.~\ref{execute}.

{\normalcolor
\begin{algorithm}[t]
\small
\caption{Training of distributed Stackelberg game algorithm based on MAPPO}
\label{A2}
\begin{algorithmic}[1]
\STATE \textbf{Initialize} actor policies $\{\pi_{\theta_m}\}_{m\in\mathcal{M}}$, centralized critic $Q_{\phi}$, and old policies $\{\pi_{\theta_m^{\rm old}}\}$ with $\theta_m^{\rm old}\leftarrow \theta_m$.
\FOR{each episode}
  \STATE Reset environment and obtain initial observations $\{o_m(1)\}_{m\in\mathcal{M}}$; initialize buffer $U\leftarrow \emptyset$.
  \FOR{$t=1$ to $T$}
    \STATE Each UAV $m$ samples $a_m(t)\sim \pi_{\theta_m^{\rm old}}(\cdot\,|\,o_m(t))$; denote ${\bf a}(t)=\{a_m(t)\}$.
    \STATE Given ${\bf a}(t)$, each MT solves its response via Algorithm~\ref{A1}; denote the follower decisions as ${\bf u}(t)$.
    \STATE Execute $\{{\bf a}(t),{\bf u}(t)\}$ in the environment; obtain rewards $\{r_m(t)\}$ and next observations $\{o_m(t+1)\}$.
    \STATE Store transition $({\bf o}(t),{\bf a}(t),{\bf r}(t),{\bf o}(t+1))$ in $U$, where ${\bf o}(t)=\{o_m(t)\}$ and ${\bf r}(t)=\{r_m(t)\}$.
    \IF{$|U|=B$ \OR $t=T$}
      \STATE Update actor policies $\{\pi_{\theta_m}\}$ using MAPPO with buffer $U$ and critic $Q_{\phi}$.
      \STATE Update critic parameters $\phi$ using buffer $U$.
      \STATE Synchronize old policies: $\theta_m^{\rm old}\leftarrow \theta_m,\ \forall m$; clear buffer $U\leftarrow\emptyset$.
    \ENDIF
  \ENDFOR
\ENDFOR
\end{algorithmic}
\end{algorithm}
}

\subsection{Complexity Analysis}
We analyze the computational complexity of the proposed algorithm from the follower-side two-stage iterative procedure and the MAPPO-based CTDE procedure. For Algorithm~\ref{A1}, let $J^{\max}$ denote the maximum number of outer best-response iterations, $I^{\max}$ denote the maximum number of inner two-stage iterations, $B^{\max}$ denote the maximum number of bisection iterations for determining the dual variables $\{\lambda_m\}$, and $\bar{M}$ denote the average number of available UAVs. In each inner iteration, updating $\gamma_k$ by \eqref{eq:gamma-star} requires evaluating the summation terms in $\gamma_k^{\rm den}$, with direct complexity $\mathcal{O}(\bar{M}^2)$ for each MT. Given $\gamma_k$, updating $\{f_{k,m}\}$ by \eqref{eq:f-star} requires computing the requested computing resources over the available UAVs and determining $\{\lambda_m\}$ to satisfy the per-UAV capacity constraints, leading to complexity $\mathcal{O}(\bar{M}^2+\bar{M}B^{\max})$ for each MT. Therefore, the complexity of Algorithm~\ref{A1} in one slot is $C_{\rm F}=\mathcal{O}(J^{\max}I^{\max}K(\bar{M}^2+\bar{M}B^{\max}))$. The actual number of iterations depends on the stopping tolerance $\epsilon$, while the worst-case complexity is upper bounded by the iteration budgets $J^{\max}$, $I^{\max}$, and $B^{\max}$. This indicates that, when the number of available UAVs and the iteration budgets are bounded, the follower-side computational cost grows linearly with the number of MTs. In practical implementation, the sparsity of the MT-UAV association variables can further reduce the cost of evaluating the summation terms.

For Algorithm~\ref{A2}, each slot involves UAV-side action sampling, follower-side response calculation, environment execution, and transition storage. Let $F_{\pi}^{\rm inf}$ denote the computational cost of one actor-network inference, and let $F_{\pi}$ and $F_Q$ denote the typical computational costs of one actor-network update and one centralized critic update, respectively. The slot-level UAV-side action sampling cost is $C_{\rm A}=\mathcal{O}(\bar{M}F_{\pi}^{\rm inf})$, while the MAPPO update cost for one buffer is $C_{\rm U}=\mathcal{O}(\bar{M}F_{\pi}+F_Q)$. Let $N_{\rm ep}$ denote the number of training episodes, $T$ denote the number of slots in each episode, and $B$ denote the buffer size for policy updates. Since Algorithm~\ref{A1} is invoked in each slot to obtain the follower-side response, the overall training complexity of Algorithm~\ref{A2} can be expressed as $\mathcal{O}(N_{\rm ep}T(C_{\rm A}+C_{\rm F})+N_{\rm ep}\lceil T/B\rceil C_{\rm U})$.
After training, the centralized critic and policy-update operations are no longer required. Each UAV only performs actor-network inference to generate its leader-side decision, while the MTs compute their follower-side responses through Algorithm~\ref{A1}. Therefore, the online execution complexity per slot is $\mathcal{O}(C_{\rm A}+C_{\rm F})$. This shows that the online computational cost is mainly determined by the number of available UAVs, the number of MTs, and the iteration budgets of the follower-side two-stage procedure. Since these factors are explicitly bounded in the algorithm implementation, the proposed algorithm can provide tractable slot-level decision-making in the considered network scale.

\section{Simulation Results and Analysis}\label{sec:simluation}

\begin{table}[t!]
\centering
\caption{Parameter Settings for Simulation}
\label{tab2}
\footnotesize
\setlength{\tabcolsep}{6pt}
\renewcommand{\arraystretch}{1.05}
\begin{tabularx}{\linewidth}{X|>{\raggedright\arraybackslash}p{0.38\linewidth}}
\hline
Parameter & Value \\
\hline
Mission length & $T=20$ \\
Slot duration & $\delta=2$ s \\
Total bandwidth & $25$ MHz \\
MT local CPU & $f_k^{\rm loc}=0.2$ GHz \\
UAV CPU budget & $f_m^{\max}=3$ GHz \\
Task input size & $J_k \in [0.1,0.5]$ Mbits \\
CPU density & $C_k=1000$ cycles/bit \\
Batch size & $B=5$ \\
Ref. channel gain & $\beta_0=-60$ dB \\
Noise power & $\sigma^2=-100$ dBm \\
MT tx power & $P_K=0.1$ W \\
UAV tx power & $P_M=1$ W \\
Capacitance coeff. & $\varphi_m=10^{-28}$ \\
Rotor tip speed & $v_{\rm tip}=110$ m/s \\
Utility weights & $\omega^{\rm W}=0.3,\ \omega^{\rm E}=0.7$ \\
PPO clip ratio & $\epsilon=0.1$ \\
Min. secure distance & $d^{\min}=5$ m \\
Algorithm accuracy & $o=0.01$ \\
Max. iterations & $J^{\max}=10$ \\
\hline
\end{tabularx}
\end{table}

We consider a dynamic UAV-enabled MEC system in a square industrial area of $[-250,250]\times[-250,250]$~m$^2$. UAVs start from four corner points, fly at altitude $H=50$ m with maximum speed $V_{\max}=25$ m/s, and maintain a minimum separation distance $d^{\min}=5$ m.
MTs are uniformly distributed and move according to a Gauss--Markov model with mean speed $0.5$ m/s and mean direction $\pi/4$, which is adopted to characterize the temporally correlated movement of MTs in a tractable manner.
Key simulation and training parameters are summarized in Table~\ref{tab2}. In particular, the default utility weights $\omega^{\rm W}=0.3$ and $\omega^{\rm E}=0.7$ are chosen to represent an energy-aware setting, since UAVs operate under limited onboard energy and their propulsion, forwarding, and computation all incur non-negligible energy consumption. Therefore, the UAV-side utility is designed to place relatively more emphasis on energy cost while still preserving the incentive for workload execution. In general, a larger $\omega^{\rm W}$ encourages more workload-oriented service decisions, while a larger $\omega^{\rm E}$ leads to more conservative decisions with stronger preference for energy saving.
To highlight the effectiveness of the proposed algorithm, the following benchmark algorithms are provided for comparison:
\begin{itemize}[leftmargin=*]
\item \textbf{No Offloading (NO)}: All MTs compute tasks locally, and UAVs hover at their takeoff locations.

\item \textbf{All Offloading (AO)}: All MTs fully offload tasks to their associated UAVs.

\item \textbf{No collaboration (NC)}: Inter-UAV task migration is disabled, and each UAV serves its associated MTs independently. This scheme serves as a no-collaboration reference in the performance comparison.
\item \textbf{Frequency Average (FA)}: Each UAV allocates computing resources equally among its associated MTs.
\item \textbf{Fixed}: Each UAV hovers near its associated MT.
\end{itemize}

\begin{table}[t] 
\centering 
\caption{Runtime evaluation of the proposed method under two representative implementation environments.}
\label{tab:runtime} 
\begingroup
\small 
\setlength{\tabcolsep}{3pt} 
\begin{tabular}{@{}c|l|c|c@{}} 
\hline 
\textbf{Env.} & \textbf{CPU/GPU} & \textbf{Total time (s)} & \textbf{Per-step (s)} \\ 
\hline
I & 
\begin{tabular}[c]{@{}l@{}} 
Intel i9-13900HX \\ 
NVIDIA RTX 5060 Laptop 
\end{tabular} 
& 0.306 & 0.0153\\ 
\hline 
II & 
\begin{tabular}[c]{@{}l@{}} 
Intel Xeon Silver 4210R \\ 
NVIDIA RTX 3090 
\end{tabular} 
& 0.275 & 0.0137 \\ 
\hline 
\end{tabular} 
\endgroup
\end{table}

To provide a practical view of the decision latency of the proposed method, we evaluate its runtime under two representative implementation environments. Environment I is configured with Python 3.10.19 and PyTorch 2.9.0, while Environment II is configured with Python 3.10.2 and PyTorch 2.5.1. As shown in Table~\ref{tab:runtime}, the proposed method completes one mission process within about 0.28--0.31 s, corresponding to an average per-step runtime of about 0.014--0.015 s. Although the two environments differ in both hardware and software configurations, the per-step runtime remains consistently below 0.02~s. This comparable millisecond-level runtime across different implementation environments supports the practicality of using the proposed method for slot-level decision making in the considered network scale.

\begin{figure}[t]
\centering
\includegraphics[width= 3.2in]{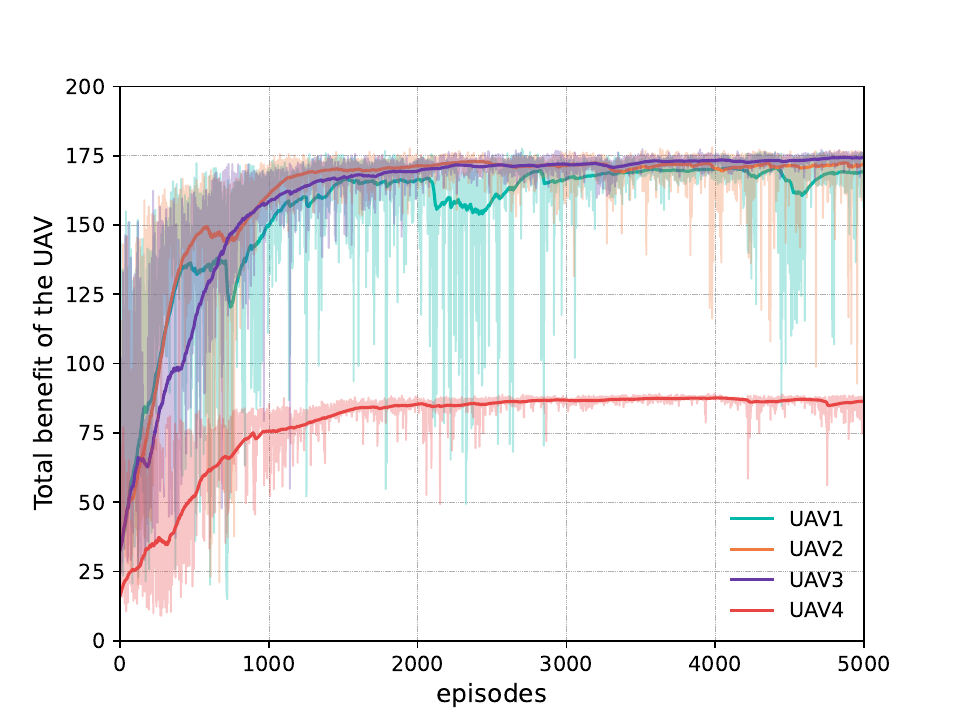}
\caption{Convergence of leader decision algorithm.}
\label{uavConver}
\end{figure}

Fig.~\ref{uavConver} shows the convergence of the leader decision algorithm. The total benefit of all UAVs remained stable after approximately 2000 training batches. Due to the fact that UAVs 1-3 provide services throughout the entire mission cycle, while UAV 4 joins the system midway through the mission and exits the system after providing a period of computing services, its total UAV benefit is approximately half that of other UAVs. 
This result further indicates that dynamic UAV participation directly affects the optimization outcome. Since UAV 4 remains available for a shorter duration, its contribution is accumulated over a smaller time span. Meanwhile, its joining and leaving behavior also changes the available service nodes for MT-UAV association and inter-UAV task migration.
Nevertheless, even when the number of available UAVs changes over time, the proposed algorithm still exhibits good convergence for different UAVs, which demonstrates its adaptability to time-varying UAV availability.

\begin{figure*}[t]
        \centering
    \begin{subfigure}[t]{0.32\textwidth}
        \centering
        \includegraphics[width=\textwidth]{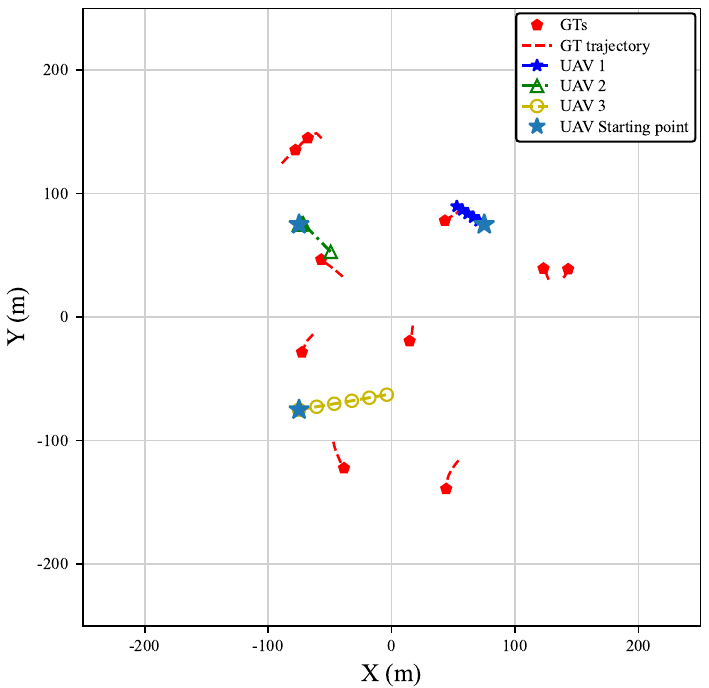}
        \caption{Phase I}
        \label{fig:traj_phase1}
    \end{subfigure}
    \hfill
    \begin{subfigure}[t]{0.32\textwidth}
        \centering
        \includegraphics[width=\textwidth]{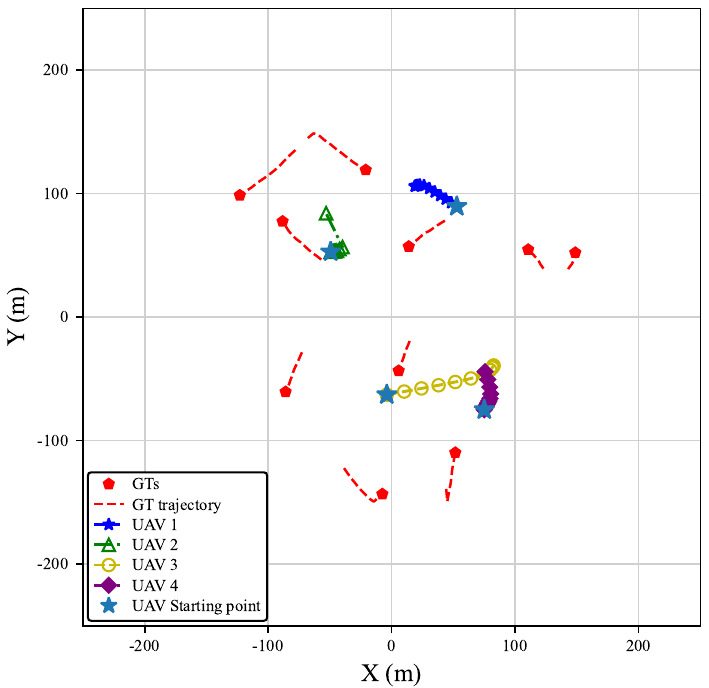}
        \caption{Phase II}
        \label{fig:traj_phase2}
    \end{subfigure}
    \hfill
    \begin{subfigure}[t]{0.32\textwidth}
        \centering
        \includegraphics[width=\textwidth]{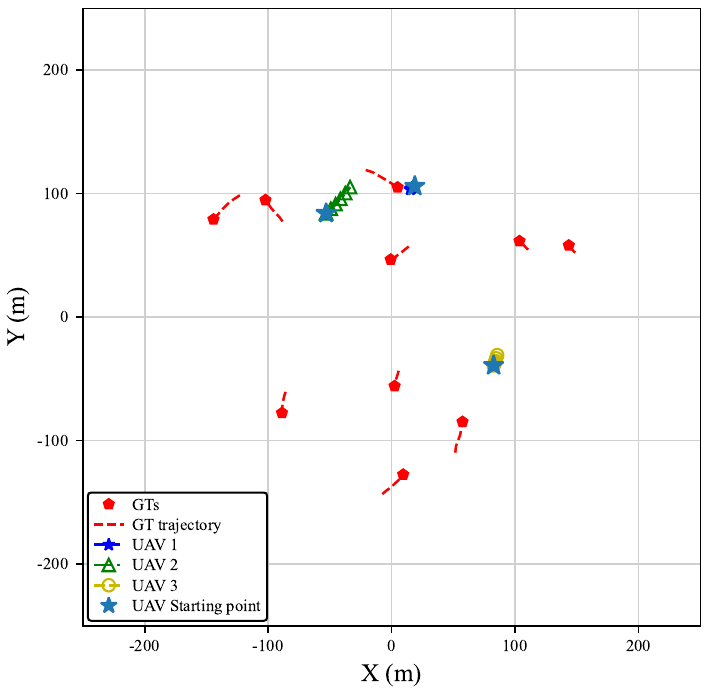}
        \caption{Phase III}
        \label{fig:traj_phase3}
    \end{subfigure}
    \caption{Representative UAV trajectories under dynamic UAV participation. The three subfigures show the three phases of one continuous mission process. In Phase I (steps 0-4), three UAVs provide service; in Phase II (steps 5-14), UAV 4 joins the system; and in Phase III (steps 15-19), UAV 4 exits the system.}
    \label{fig:traj_dynamic}
    \vspace{-18pt}
\end{figure*}

Fig.~\ref{fig:traj_dynamic} shows UAV trajectories in the three phases of one continuous mission process under dynamic UAV participation, illustrating how the UAV spatial deployment is reconfigured as the available UAV set changes over time.
In Phase I, the three UAVs exhibit a relatively stable spatial deployment, and their trajectories mainly involve local adjustments around their service regions. 
After UAV 4 joins in Phase II, the trajectory reconfiguration appears in the lower service region, where UAV 3 and UAV 4 move within a nearby area, suggesting that the newly available UAV mainly enhances local service flexibility and cooperative support in that region, while the other UAVs largely remain in their original service areas. 
After UAV 4 exits in Phase III, the remaining UAVs further adjust their trajectories and gradually recover a relatively separated spatial deployment pattern. 
Overall, these phase-wise trajectory changes suggest that the proposed framework can maintain continuous and coordinated UAV service behavior under time-varying UAV availability.

\begin{figure}[t]
\centering
\includegraphics[width= 3.2in]{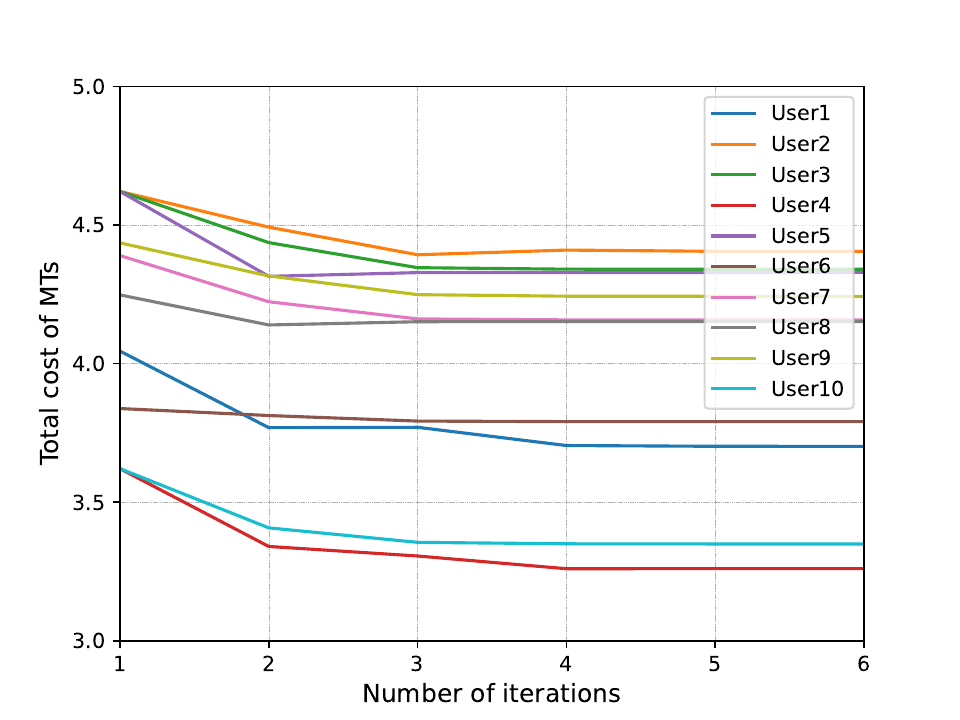}
\caption{Convergence of follower decision algorithm.}
\label{userConver}
\end{figure}

Fig.~\ref{userConver} shows the convergence of the follower decision algorithm. The total overhead of each of the 10 MTs remained stable after about 4 iterations of the proposed algorithm. Due to the analytical solution obtained by analyzing the KKT conditions, the follower decision algorithm has a fast overall iteration, which can meet the requirements of time-sensitive tasks and achieve online computing.

\begin{figure}[!t]
\centering
\includegraphics[width= 3.2in]{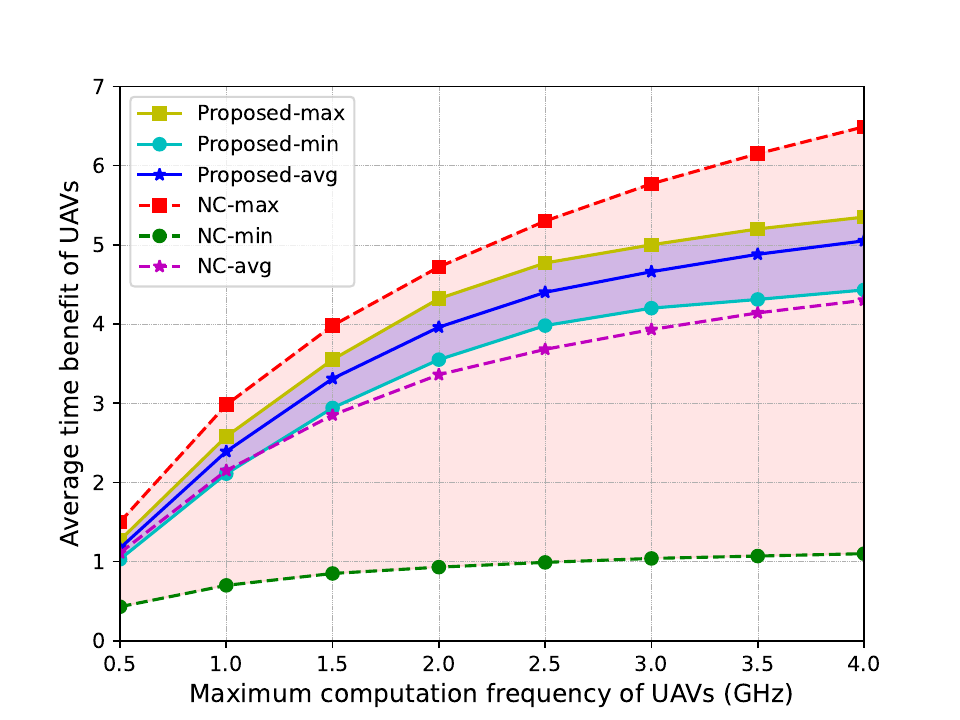}
\caption{Average time benefit of UAVs at different maximum computing frequencies.}
\label{uav_cpu}
\end{figure}

Fig.~\ref{uav_cpu} shows the average time efficiency of UAVs at different maximum computing frequencies. As the maximum computing frequency of UAVs increases, the benefits obtained by selling computing resources gradually increase, and the average time efficiency of UAVs gradually increases. However, due to the limited energy consumption caused by task computing and the limited demand for computing resources from MTs, the increase in average time efficiency of UAVs is gradually decreasing. In the absence of collaborative work between UAVs, the uneven distribution of MT locations leads to different UAV loads, resulting in extreme values of UAV time average efficiency. 
Although the maximum time-averaged benefit of the NC algorithm for UAVs is relatively higher compared to the proposed algorithm, the average benefit of the NC algorithm is lower due to the small number of UAVs connected to MTs, which indicates that enabling inter-UAV collaboration is beneficial to improving UAV resource utilization under the considered setting.

\begin{figure}[!t]
\centering
\includegraphics[width= 3.2in]{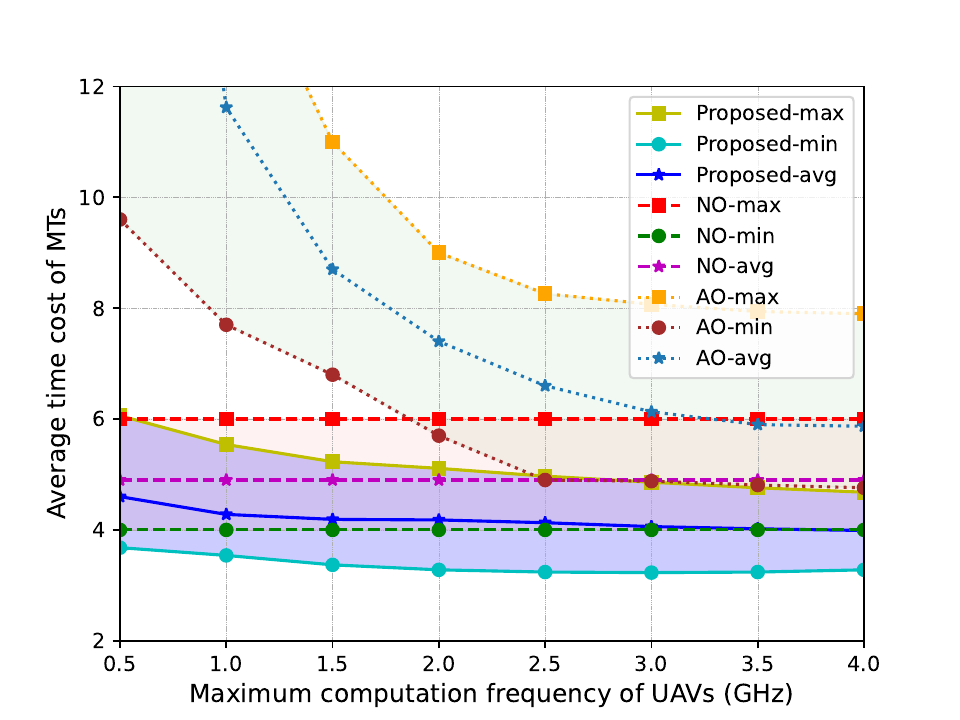}
\caption{Average MT time overhead at different maximum computing frequencies for UAVs.}
\label{user_cpu}
\end{figure}

The average time overhead of MTs at different maximum computing frequencies for UAVs is given in Fig.~\ref{user_cpu}. With the increase of the maximum computing frequency of UAVs, the average time overhead of the proposed algorithm and AO corresponding to MTs is gradually decreasing. MTs can purchase more computing resources from UAVs, and the computing time for offloading tasks is reduced. The limited computing tasks of MTs and the pricing of UAV computing frequency mean that MTs have limited demand for computing resources, so the reduction in average time overhead of MTs is gradually decreasing. Due to the fact that all computing tasks corresponding to NO are executed locally and are not affected by the frequency of UAV computing. The proposed algorithm has better performance compared to NO and AO, indicating that partial offloading computing tasks can efficiently utilize both local resources of MTs and UAV computing resources simultaneously.

\begin{figure}[!t]
\centering
\includegraphics[width= 3.2in]{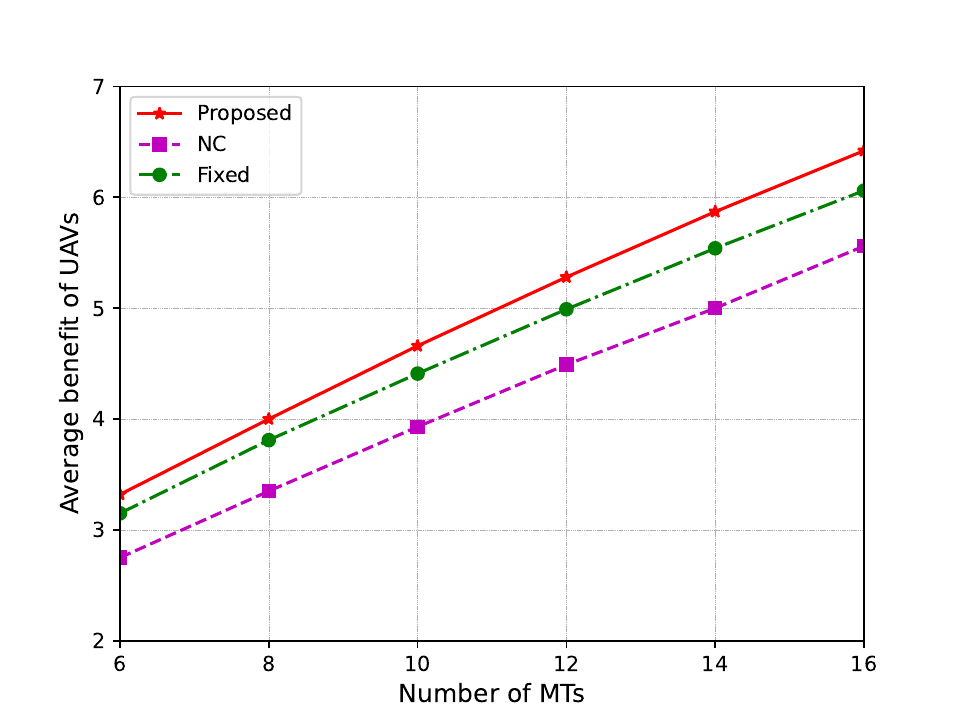}
\caption{Average benefits of UAVs at different numbers of MTs.}
\label{uav_user}
\end{figure}

In Fig.~\ref{uav_user}, we show the average benefits of UAVs under different numbers of MTs. The average benefits of UAVs increase with the increase of the number of MTs, and the unit computing task volume is directly proportional to the UAV benefits, thus showing a linear growth overall. Fixed UAV positioning can reduce the flight energy consumption of UAVs to a certain extent, but due to the inability to actively approach MTs to reduce communication distance, it leads to an increase in communication energy consumption and communication latency, resulting in a lower overall average efficiency of UAVs. 
The NC algorithm performs the worst, which further highlights the importance of coordinated trajectory planning and inter-UAV collaboration for improving the average efficiency of~UAVs.

\begin{figure}[!t]
\centering
\includegraphics[width= 3.2in]{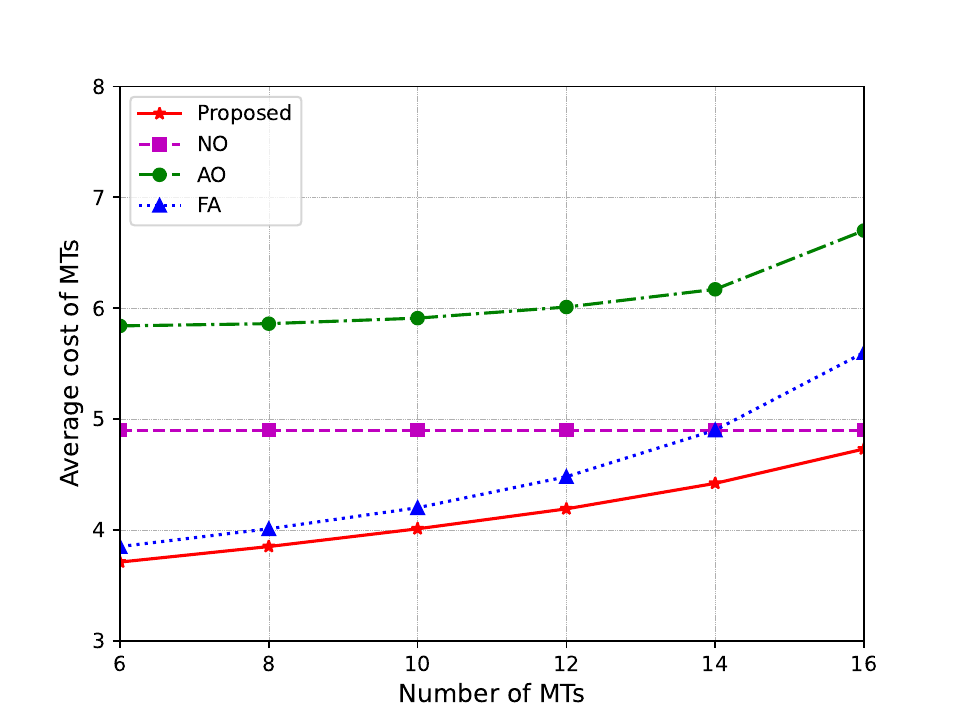}
\caption{Average MT overhead at different numbers of MTs.}
\label{user_user}
\end{figure}

We investigate the average overhead of MTs under different numbers of MTs in Fig.~\ref{user_user}. Due to the fact that the calculation tasks corresponding to the NO algorithm are all executed locally on MTs, the MT overhead has little impact on NO after averaging the number of MTs. Under the constraint of limited computing resources in UAVs, as the number of MTs increases, the computing resources obtained by each MT decrease, and the task computing delay increases accordingly. Therefore, the average overhead of MTs gradually increases. When the difference in task volume between MTs is not significant, evenly distributing UAV computing resources is a relatively effective method, but the performance of the proposed algorithm is still relatively poor.

\begin{figure}[!t]
\centering
\includegraphics[width= 3.2in]{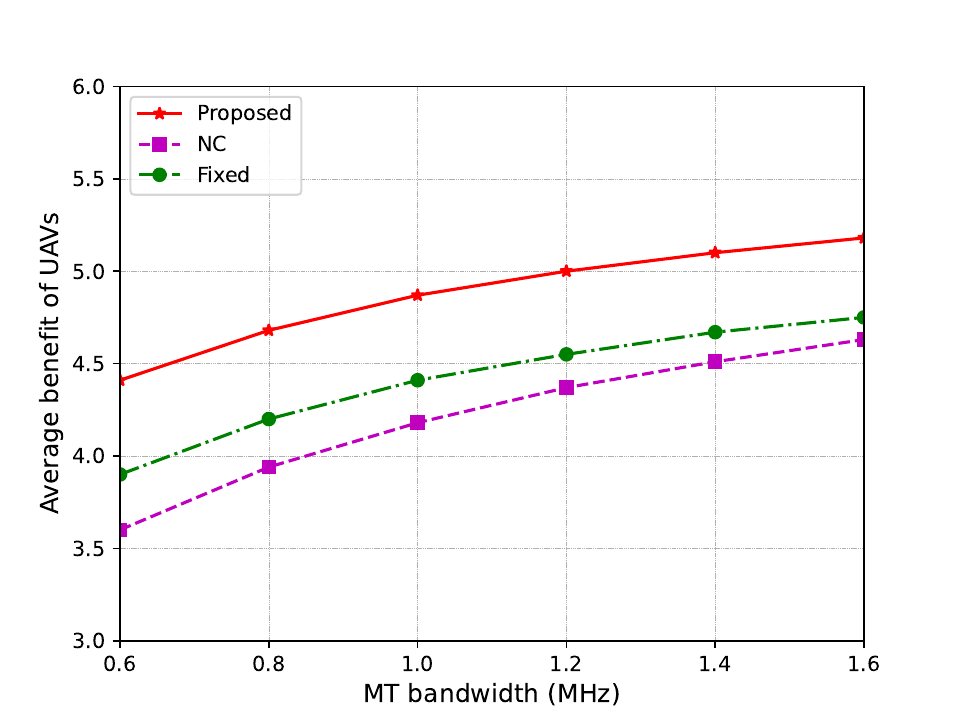}
\caption{Average benefits of UAVs at different MT bandwidths.}
\label{uav_userBand}
\end{figure}

\begin{figure}[!t]
\centering
\includegraphics[width= 3.2in]{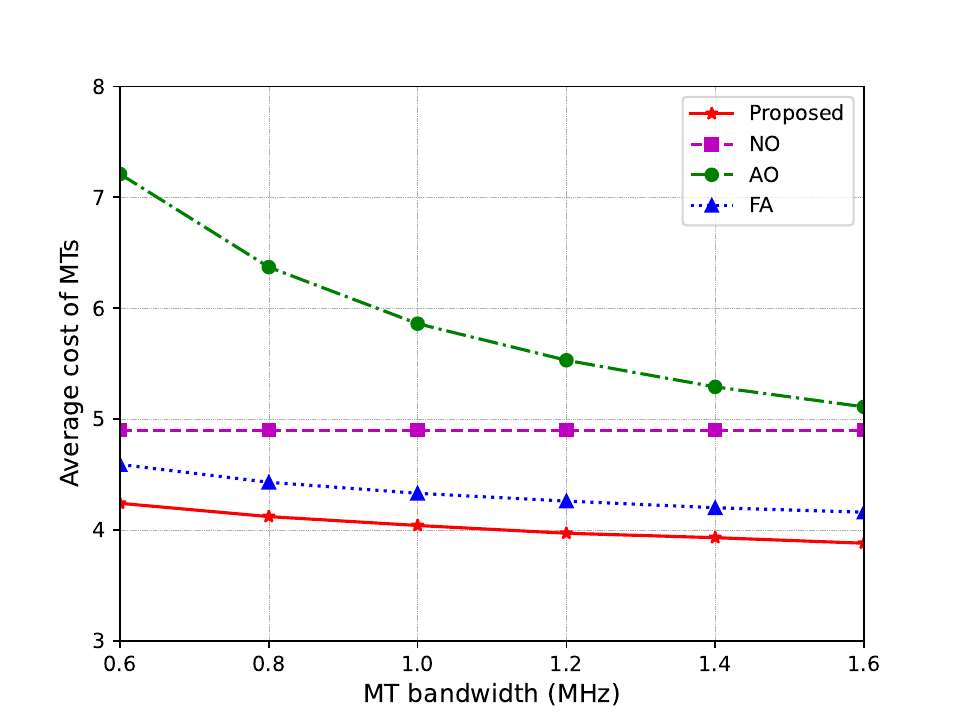}
\caption{Average MT overhead at different MT bandwidths.}
\label{user_userBand}
\end{figure}

We analyze the average benefits of UAVs under different MT bandwidths in Fig.~\ref{uav_userBand}. As the bandwidth of MTs increases, the average efficiency of UAVs gradually increases due to the decrease in communication latency caused by high bandwidth, thereby reducing communication energy consumption. It can be seen that the increase in average efficiency of UAVs is gradually slowing down because there is a logarithmic relationship between MT bandwidth and energy consumption. This also means that increasing MT bandwidth within a certain range is feasible, but when the bandwidth is too large, the system benefits brought by increasing bandwidth are very small. The proposed algorithm still maintains good performance compared to the benchmark algorithm.

Finally, we show the average overhead of MTs under different bandwidths in Fig.~\ref{user_userBand}. With the increase of MT bandwidth, the overall average overhead of MTs is gradually decreasing. Due to the fact that the NO algorithm does not offload any computing tasks and there is no scenario for communication with UAVs, it is not affected by the bandwidth of MTs. The average overhead of MTs corresponding to other algorithms gradually decreases with the increase of MT bandwidth, and the decrease rate gradually decreases. The equal distribution of UAV computing resources still has good performance and can serve as an alternative to the proposed algorithm to a certain extent. But the proposed algorithm still performs well under different conditions.

\section{Conclusion}\label{sec:conclusion}
In this paper, we proposed a dynamic distributed MEC architecture to support collaborative computing among multiple UAVs in scenarios with mobile MTs and randomly generated computing tasks, enabling the dynamic joining and leaving of UAVs.  
MTs were allowed to partially offload their computing tasks to connected UAVs, while overloaded UAVs could further redistribute tasks through collaborative migration.  
We formulated a Stackelberg game-based optimization model to maximize UAV benefits by jointly optimizing UAV trajectories, inter-UAV task migration ratios, UAV access decisions, and computing resource pricing and solved it using a MAPPO-based algorithm.  
Simultaneously, MT overhead was minimized by jointly optimizing computing resource allocation and task offloading ratios, with a two-stage iterative method adopted to solve the follower subproblem.
Simulation results demonstrated that the proposed algorithms achieved stable convergence and significantly outperformed benchmark schemes under varying computing capacities, MT densities, and bandwidth conditions.
Future work may extend the current model by incorporating more adaptive bandwidth allocation mechanisms to further improve spectrum utilization and provide a more comprehensive characterization of uplink resource management in distributed multi-UAV MEC systems.
In addition, practical deployment and validation on real UAV platforms will be investigated by integrating measured wireless links, onboard computing constraints, flight-control limitations, and online task offloading decisions into a closed-loop experimental framework, which can further evaluate the robustness and practical applicability of the proposed method under real-world operating conditions.

{
\bibliographystyle{IEEEtran}
\bibliography{mybib}
}

\begin{IEEEbiography}[{\includegraphics[width=1in,height=1.25in,clip,keepaspectratio]{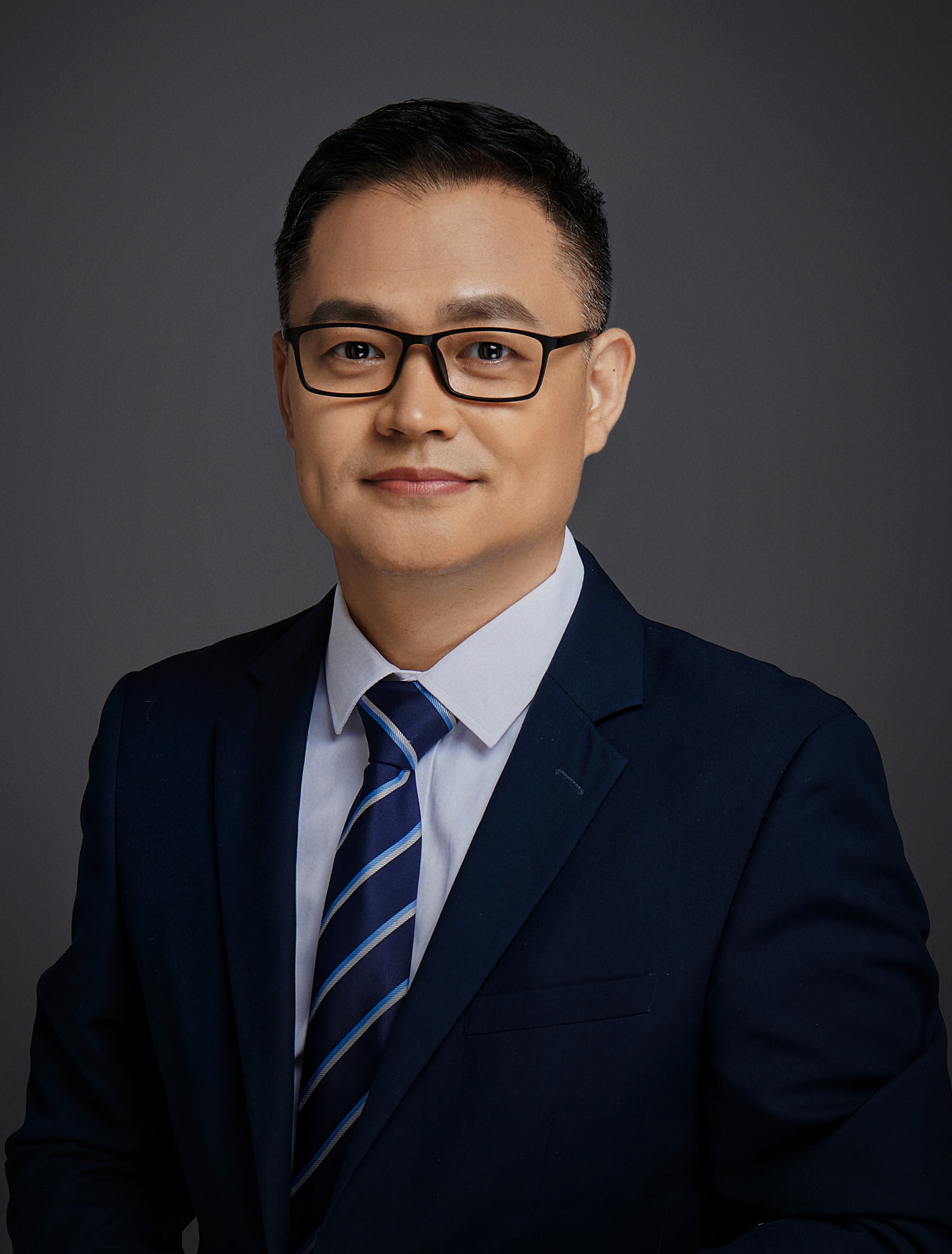}}]{Tiankui Zhang} (M'10-SM'15) received the Ph.D. degree in Information and Communication Engineering and B.S. degree in Communication Engineering from Beijing University of Posts and Telecommunications (BUPT), China, in 2008 and 2003, respectively. Currently, he is a Professor in School of Information and Communication Engineering at BUPT. His research interests include artificial intelligence enabling wireless networks, UAV communications in 5G and beyond networks, intelligent mobile edge computing, signal processing for wireless communications. He had published more than 240 papers including journal papers on IEEE Journal on Selected Areas in Communications, IEEE Transactions on Communications, etc., and conference papers, such as IEEE GLOBECOM and IEEE ICC. He is the co-recipient of the Best Paper Award in IEEE GLOBECOM 2022, the Best Paper Award in IEEE APCC 2011. He has served as a TPC Member for many IEEE conferences, such as GLOBECOM and PIMRC, the Technical Program Committee Chair for AiCON 2021. He has served as the guest editor for the special issue "Future Network Architecture and Key Technologies" of the Journal of Beijing University of Posts and Telecommunications, the special issue "Recent Advances in UAV Communications and Networks" of Sensors.
\end{IEEEbiography}

\begin{IEEEbiography}[{\includegraphics[width=1in,height=1.25in,clip,keepaspectratio]{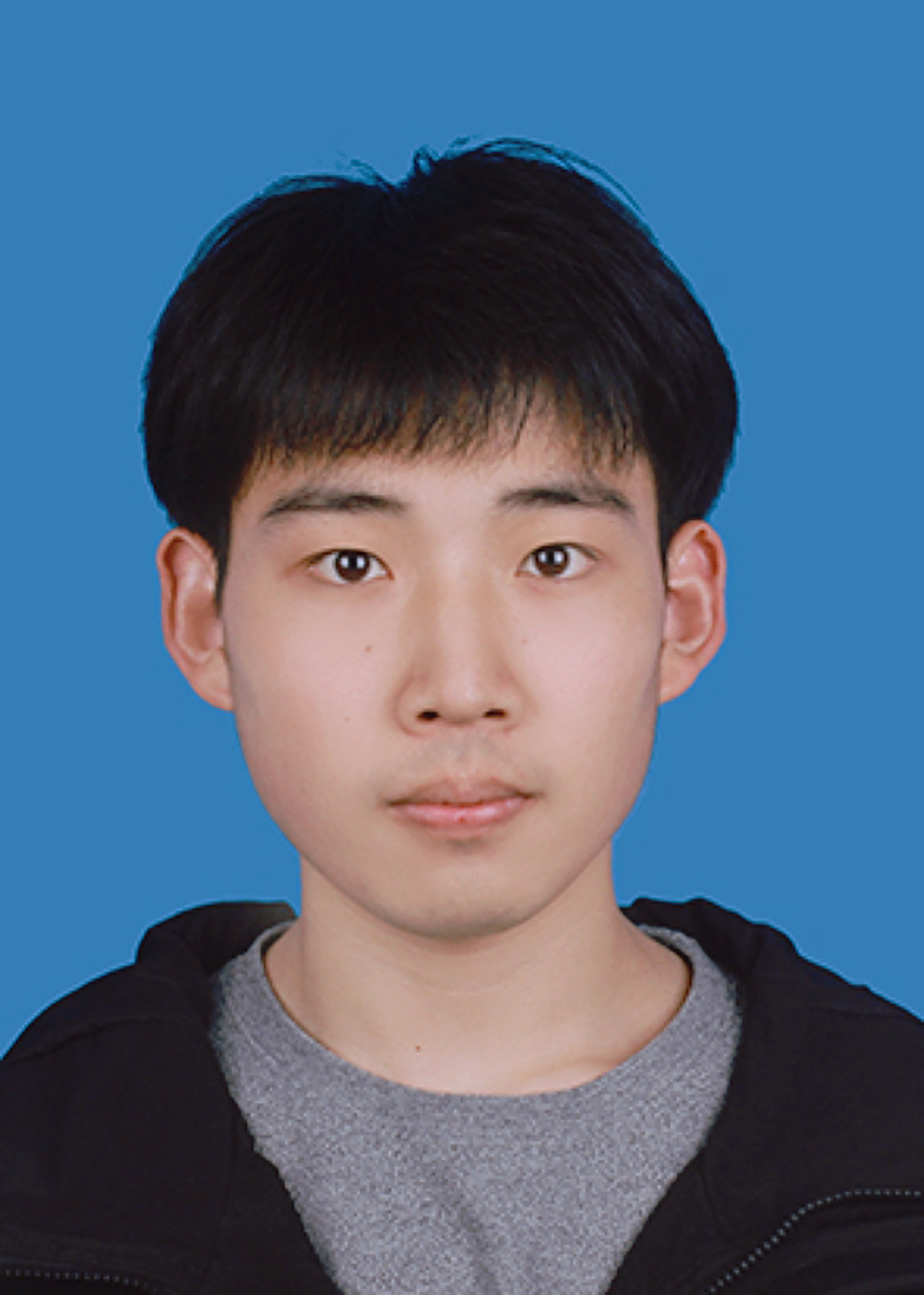}}]{Wenlong Xu} received the B.S. degree from China Agricultural University (CAU), Beijing, China, in 2022, and the M.S. degree from, Beijing University of Posts and Telecommunications (BUPT), Beijing, China, in 2025. His current research interests include UAV trajectory planning and resource allocation in mobile edge computing.
\end{IEEEbiography}

\begin{IEEEbiography}[{\includegraphics[width=1in,height=1.25in,clip,keepaspectratio]{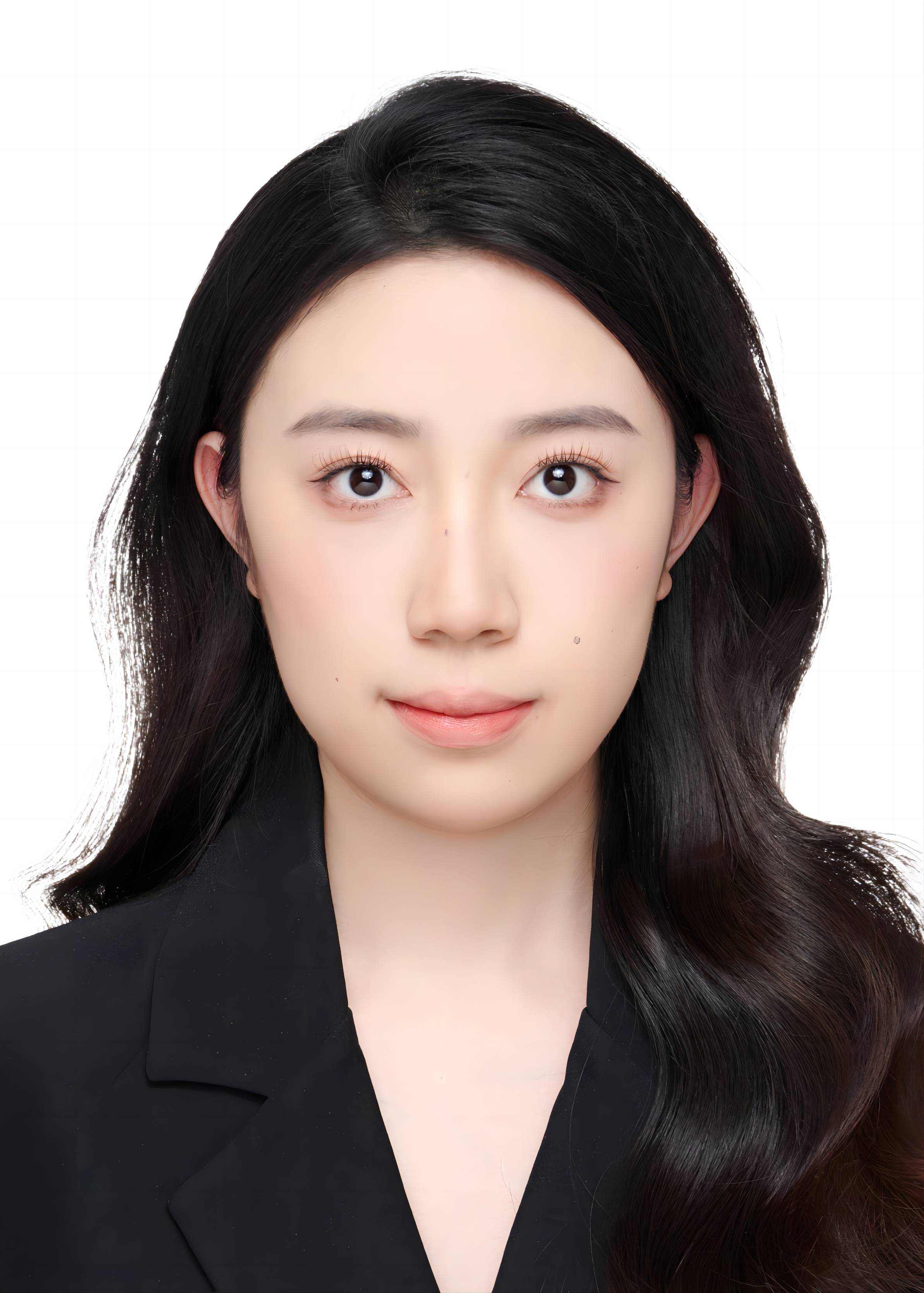}}]{Tianyi Shi} (Graduate Student Member, IEEE) received the B.S. degree in Communication Engineering from Harbin Institute of Technology, Weihai, China, in 2021. She is currently pursuing the Ph.D. degree in Information and Communication Engineering at Beijing University of Posts and Telecommunications, Beijing, China. She was a visiting Ph.D. student at Queen's University Belfast, U.K., from September 2025 to August 2026. Her research interests include mobile edge computing, edge intelligence, and AI for wireless networks.
\end{IEEEbiography}

\begin{IEEEbiography}[{\includegraphics[width=1in,height=1.25in,clip,keepaspectratio]{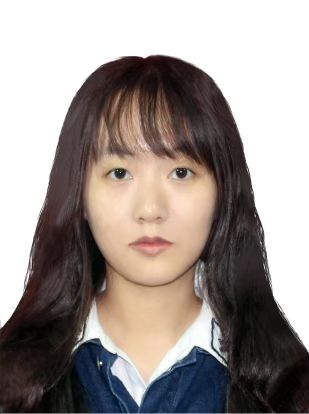}}]{Xiaoxia Xu} (Member, IEEE) received the B.Eng. and Ph.D. degrees from Wuhan University, China, in 2017 and 2023, respectively. From 2021 to 2022, she was a Visiting Student with the Queen Mary University of London (QMUL), London, U.K. She is currently a Post-Doctoral Researcher with the School of Electronic Engineering and Computer Science, QMUL. Her current research interests include millimetre-wave/terahertz communications, flexible-antenna techniques, next generation multiple access, AI for B5G/6G, and edge intelligence. For more information \url{https://xiaoxiaxusummer.github.io/}
\end{IEEEbiography}

\begin{IEEEbiography}[{\includegraphics[width=1in,height=1.25in,clip,keepaspectratio]{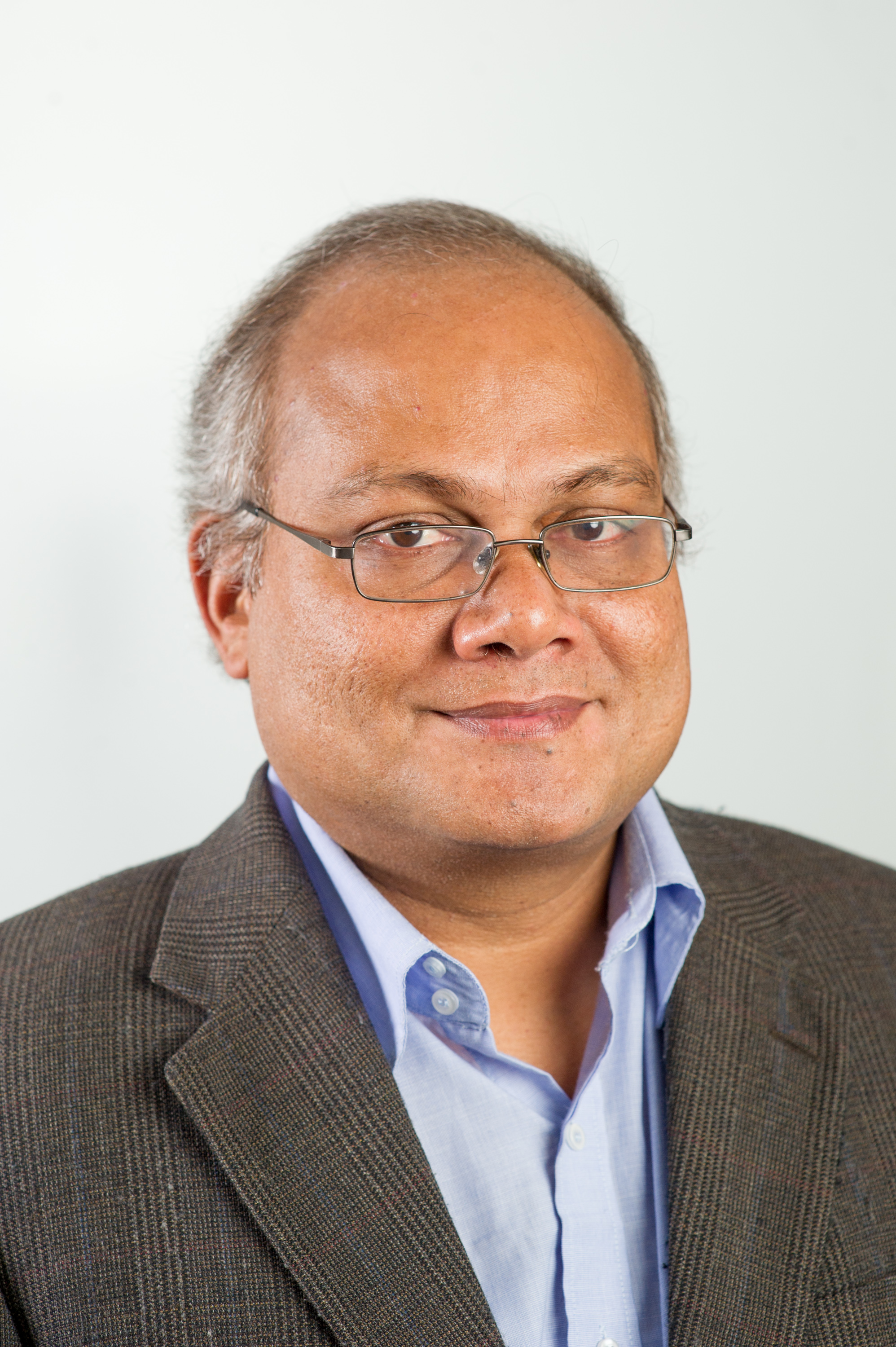}}]{Arumugam Nallanathan} (Fellow, IEEE) has been a Professor in wireless communications and the Founding Head of the Communication Systems Research (CSR) Group, School of Electronic Engineering and Computer Science, Queen Mary University of London, since September 2017. He was with the Department of Informatics at King’s College London from December 2007 to August 2017, where he was Professor of Wireless Communications from April 2013 to August 2017 and a Visiting Professor from September 2017 till August 2020. He was an Assistant Professor in the Department of Electrical and Computer Engineering, National University of Singapore from August 2000 to December 2007. His research interests include Artificial Intelligence for Wireless Systems, Beyond 5G Wireless Networks and Internet of Things (IoT). He published more than 700 technical papers in scientific journals and international conferences. He is a co-recipient of the Best Paper Awards presented at the IEEE International Conference on Communications 2016 (ICC'2016), IEEE Global Communications Conference 2017 (GLOBECOM'2017) and IEEE Vehicular Technology Conference 2018 (VTC'2018). He is also a co-recipient of IEEE Communications Society Leonard G. Abraham Prize in 2022. He is an IEEE Distinguished Lecturer. He has been selected as a Web of Science Highly Cited Researcher in 2016, and 2022-2024.  

He was a Senior Editor for IEEE Wireless Communications Letters, an Editor for IEEE Transactions on Wireless Communications, IEEE Transactions on Communications, IEEE Transactions on Vehicular Technology and IEEE Signal Processing Letters. He served as a Guest Editor for numerous special issues of IEEE Journal on Selected Areas in Communications (JSAC). He served as the Chair for the Signal Processing and Communication Electronics (SPCE) Technical Committee of IEEE Communications Society and Technical Program Chair and member of Technical Program Committees in numerous IEEE conferences. He received the IEEE Communications Society SPCE outstanding service award 2012 and IEEE Communications Society RCC outstanding service award 2014.
\end{IEEEbiography}

\end{document}